\documentclass[a4paper,fleqn,usenatbib]{mnras}

\usepackage{txfonts}

\usepackage[T1]{fontenc}
\usepackage{ae,aecompl}

\usepackage{graphicx}	% Including figure files
\usepackage{amstext}
\usepackage{amssymb}	% Extra maths symbols

\newcommand{\HI}{$\rm{H\,{\sevensize I}}$}

\newcommand{\lya}{Ly$\alpha$}

\newcommand{\kms}{$\rm km~s^{-1}$}

\newcommand{\MgII}{\mbox{Mg\,{\sc ii}}}
\newcommand{\MgI}{\mbox{Mg\,{\sc i}}}
\newcommand{\FeII}{\mbox{Fe\,{\sc ii}}}
\newcommand{\CaII}{\mbox{Ca\,{\sc ii}}}
\newcommand{\MnII}{\mbox{Mn\,{\sc ii}}}
\newcommand{\CrII}{\mbox{Cr\,{\sc ii}}}
\newcommand{\AlIII}{\mbox{Al\,{\sc iii}}}

\newcommand{\qsoone}{QSO-SE}
\newcommand{\qsotwo}{QSO-NW}

\title[MUDF. II. The gaseous properties of galaxy groups]{The MUSE Ultra Deep Field (MUDF). II. Survey design and the gaseous properties of galaxy groups at $0.5<z<1.5$}

\author[Fossati et al.]{M. Fossati$^{1,2}$\thanks{E-mail: matteo.fossati@durham.ac.uk}, 
M. Fumagalli$^{1,2,3}$, 
E.K. Lofthouse$^{1,2}$, 
V. D'Odorico$^{4,5}$,
E. Lusso$^{6,7}$,\and
S. Cantalupo$^8$,
R.J. Cooke$^{1,2}$,
S. Cristiani$^4$,
F. Haardt$^{9,10}$,
S.L. Morris$^{2,11}$,\and
C. Peroux$^{12,13}$,
L.J. Prichard$^{14}$,
M. Rafelski$^{14,15}$,
I. Smail$^2$,
T. Theuns$^{1}$
\\
  $^{1}$Institute for Computational Cosmology, Durham University, South Road, Durham, DH1 3LE, UK \\
  $^{2}$Centre for Extragalactic Astronomy, Durham University, South Road, Durham, DH1 3LE, UK \\
  $^{3}$Dipartimento di Fisica G. Occhialini, Universit\`a degli Studi di Milano-Bicocca, Piazza della Scienza 3, 20126 Milano, Italy \\
  $^{4}$INAF, Osservatorio Astronomico di Trieste, Via G. B. Tiepolo 11, 34143 Trieste, Italy \\
  $^{5}$Scuola Normale Superiore, Piazza dei Cavalieri 7, 56126 Pisa, Italy \\
  $^{6}$Dipartimento di Fisica e Astronomia, Universit\`a di Firenze, via G. Sansone 1, 50019 Sesto Fiorentino, Firenze, Italy\\
  $^{7}$INAF--Osservatorio Astrofisico di Arcetri, 50125 Firenze, Italy\\
  $^{8}$Department of Physics, ETH Zurich, Wolgang-Pauli-Strasse 27, 8093, Zurich \\
  $^{9}$DiSAT, Universit\`a degli Studi dell'Insubria, Via Valleggio 11, I-22100 Como, Italy \\
  $^{10}$INFN, Sezione di Milano-Bicocca, Piazza della Scienza 3, I-20126 Milano, Italy \\
  $^{11}$Centre for Advanced Instrumentation, Durham University, South Road, Durham DH1 3LE, UK \\
  $^{12}$European Southern Observatory (ESO), Karl-Schwarzschild-Str.2, D-85748 Garching b. M{\"u}nchen, Germany \\
  $^{13}$Aix Marseille Universit{\'e}, CNRS, LAM (Laboratoire d'Astrophysique de Marseille) UMR 7326, 13388, Marseille, France \\
  $^{14}$Space Telescope Science Institute, 3700 San Martin Drive, Baltimore, MD 21218, USA \\
  $^{15}$Department of Physics and Astronomy, Johns Hopkins University, Baltimore, MD 21218, USA \\
 }

\date{Accepted 2019 September 17. Received 2019 September 14; in original form 2019 July 30.}

\pubyear{\the\year}

\begin{document}
\label{firstpage}
\pagerange{\pageref{firstpage}--\pageref{lastpage}}
\maketitle

% Abstract of the paper
\begin{abstract}
We present the goals, design, and first results of the MUSE Ultra Deep Field (MUDF) survey, a large programme using the Multi Unit Spectroscopic Explorer (MUSE) instrument at the ESO Very Large Telescope. The MUDF survey is collecting $\approx 150$ hours on-source of integral field optical spectroscopy in a $1.5\times1.2$ arcmin$^2$ region which hosts several astrophysical structures along the line of sight, including two bright $z\approx 3.2$ quasars with close separation ($\approx 500$ kpc). 
Following the description of the data reduction procedures, we present the analysis of the galaxy environment and gaseous properties of seven groups detected at redshifts $0.5<z<1.5$, spanning a large dynamic range in halo mass, $\log(M_h/\rm{M_\odot}) \approx 11 - 13.5$. For four of the groups, we find associated \MgII\ absorbers tracing cool gas in high-resolution spectroscopy of the two quasars, including one case of correlated absorption in both sightlines at distance $\approx 480$ kpc.
The absorption strength associated with the groups is higher than what has been reported for more isolated galaxies of comparable mass and impact parameters. We do not find evidence for widespread cool gas { giving rise to strong absorption} within these groups. Combining these results with the distribution of neutral and ionised gas seen in emission in lower-redshift groups, we conclude that gravitational interactions in the group environment strip gas from the galaxy haloes into the intragroup medium, boosting the cross section of cool gas and leading to the high fraction of strong \MgII\ absorbers that we detect.
\end{abstract}

% Select between one and six entries from the list of approved keywords.
% Don't make up new ones.

%%%%%%%%%%%%%%%%%%%%%%%%%%%%%%%%%%%%%%%%%%%%%%%%%%

%%%%%%%%%%%%%%%%% BODY OF PAPER %%%%%%%%%%%%%%%%%%

\section{Introduction}
In the last few decades, observations from ground and space based telescopes coupled to theoretical models have built the Lambda cold dark matter ($\Lambda$CDM) model. In this framework galaxies form within dark matter haloes that grow hierarchically \citep[e.g.][]{Gunn72, White78, Perlmutter99}.
Several, and often competing, processes then lead to the growth, evolution and final fate of galaxies inside these haloes, giving rise to the diverse morphology, colours, and structural properties of present-day galaxies. One of the major tasks of modern theories of galaxy formation is thus to describe in detail the physical processes that underpin this observed evolution.

Galaxies grow both in mass and size \citep{Muzzin13,van-der-Wel14} by acquiring gas either through cooling of a hot gas halo, or via cold gas streams fed by cosmic filaments \citep[e.g][]{Keres05,Dekel06,van-de-Voort11}. The process of star formation converts the gas into stars and it is tightly related to the net balance of gas accretion and ejection \citep{Bouche10, Dave12, Lilly13, Sharma19}. Feedback processes related to stellar winds, supernova explosions, and active galactic nuclei eject gas back into the { haloes surrounding galaxies: the} circumgalactic medium (CGM). While some of this gas falls back on the galaxy in a fountain-like mode \citep{Fraternali08}, a fraction of it can leave the halo of galaxies contributing to the metal enrichment of the intergalactic medium \citep[IGM, e.g][]{Dekel86,Schaye03,Springel05,Oppenheimer10}. Within this framework, it becomes crucial for an effective theory of galaxy evolution to understand how inflows and outflows interact and coexist within the CGM \citep[e.g.][]{Steidel10,Tumlinson17}, that is the gaseous component surrounding galaxies.

At the same time, galaxies do not evolve in isolation, but are accreted onto more massive structures, comprising groups or clusters of galaxies. Indeed, it has long been known that galaxies in dense environments have different properties than those in less dense environments, both in terms of gas content and structural properties \citep[e.g.][]{Oemler74,Dressler80,Giovanelli85,Balogh04,Peng10,Fossati17}. Environmental processes, triggered by the interactions between galaxies themselves with the hot medium in massive haloes { and with the IGM} further regulate the gas supply of galaxies \citep[e.g.][]{Boselli06}, leading to { a different evolution compared to more isolated galaxies.}  

Significant progress has been made in understanding the gas-galaxy co-evolution since the advent of large spectroscopic surveys of galaxies at $z\lesssim 1$ (e.g. SDSS; \citealt{York00} or GAMA; \citealt{Driver11}). These galaxy surveys have been complemented by spectroscopic observations of quasars allowing for detailed studies of the CGM in absorption as a function of galaxy properties (including mass, star-formation rates), and their environment \citep[e.g.][]{Prochaska11,Tumlinson13,Tumlinson17,Stocke13,Bordoloi14,Tejos14,Finn16,Kauffmann17}. 
These studies reveal the ubiquitous presence of a multiphase, enriched, and kinematically-complex CGM surrounding every galaxy \citep{Werk14, Werk16}. 
This picture of a multiphase CGM has been extended to $z>1$ thanks to extensive observational campaigns with multi-object spectrographs on 8-10m telescopes  \citep[e.g.][]{Steidel10,Rubin10,Crighton11,Rudie12,Turner14,Tummuangpak14,Turner17}. Despite these advancements, however, our view of the CGM at early cosmic epochs has been mostly limited to star-forming galaxies at the bright end of the UV luminosity function, and to scales of $\approx 0.1-1~\rm Mpc$ around galaxies, due to the difficulty in obtaining spectroscopy of samples of objects at small projected separations from the quasars using traditional multi-objects spectrographs. 

These limitations have recently been lifted by integral field spectrographs which have been deployed at the largest observing facilities, including the Multi Unit Spectroscopic Explorer \citep[MUSE;][]{Bacon10} at the ESO Very Large Telescope, and the Keck Cosmic Web Imager \citep[KCWI;][]{Morrissey18} at the Keck Observatory. In particular, thanks to its large 1 arcmin$^2$ field of view, its extended wavelength coverage in the optical, and its exquisite sensitivity, MUSE has become the ideal instrument for studies of the CGM of galaxies in quasar fields \citep[e.g.][]{Schroetter16,Fumagalli16, Fumagalli17a, Peroux17, Bielby17a, Klitsch18, Chen19a}.

Leveraging the unique features of this instrument, we have designed an observational campaign to acquire very deep MUSE observations in the field centred at 21$^{h}$:42$^{m}$:24$^{s}$,$-$44$^{\circ}$:19$^{m}$:48$^{s}$
(hereafter the MUSE Ultra Deep Field or MUDF). This field stands out for its rare property of hosting two bright quasars at $z\approx 3.22$ that probe the IGM and CGM of intervening galaxies with two sightlines $\approx 60$ arcsec apart. Another quasar lies at the same redshift at $\approx 8$ arcmin separation, making this system a quasar triplet \citep{Francis93}. Upon its completion, this programme will acquire $\approx 200~$hours of MUSE data (corresponding to $\approx 150$ hours on source) in a $1.5 \times 1.2$ arcmin$^2$ region around the two quasars (ESO PID 1100.A$-$0528). This programme is complemented by deep high-resolution spectroscopy of the quasars using the UV and Visual Echelle Spectrograph \citep[UVES][]{Dekker00} at the VLT (ESO PIDs 65.O$-$0299, 68.A$-$0216, 69.A$-$0204, 102.A$-$0194), %(see Table \ref{tab:uves} for PIDs), 
and by the deepest spectroscopic survey (90 orbits in a single field) in the near infrared using the Wide Field Camera 3 instrument on board the {\it Hubble Space Telescope}, together with deep 8-orbit near UV imaging ({\it HST} PIDs 15637 and 15968).  

These combined datasets will enable us to achieve several goals. First and foremost, we will connect the presence of gas in the CGM of galaxies with their properties and their environment from $z\approx 3$ to the present day. Without a pre-selection for UV and optically bright sources, we will have a unique vantage point on the low mass galaxy population up to $z\sim3$. Furthermore, the MUDF hosts notable structures as a function of redshift. For instance, a correlated strong \HI\ absorber detected in both sightlines at $z\approx 3$ hints at an extended structure running across the field of view \citep{DOdorico02}. Moreover, the presence of a quasar pair is suggestive of an overdense region at $z\approx 3.22$, which is predicted to lie at the intersection of filaments in the cosmic web. Indeed, in the first paper of this series, \citet{Lusso19} studied the morphology of the giant \lya\ nebulae surrounding the quasars, finding an elongation of the ionized gas along the line connecting the two quasars. In the future, once we have the full dataset, we will search for the presence of ionized gas in this putative filament. The depth of the observations will also provide spectra of exquisite quality for a few hundred objects. Thanks to our multiwavelength dataset from the near-UV to the near-IR, we will study the properties and structure of low-mass galaxies across a large fraction of cosmic time. 

 %Most of these goals require the depth given by the full dataset, which is gradually been collected.
 In this paper, we present the survey design and the details of the MUSE observations and data reduction. 
 As a first application we focus on the connection of enriched cool gas ($T\sim 10^4$ K), as traced by the \MgII\ $\lambda \lambda\ 2796,2803 \AA$ absorption doublet in the quasar spectra, with the galaxy population and its environment at $0.5<z<1.5$. 
 Even though the full MUSE dataset is still being collected, the observations available to date already provide an excellent dataset for an accurate reconstruction of the local galaxy environment, and of the physical properties of galaxies at $z \lesssim 1.5$. Several studies have used the \MgII\ doublet to trace gas with similar column densities to that detected through 21-cm atomic hydrogen observations \citep[e.g.][]{Bergeron86,Kacprzak08,Steidel94,Chen08,Chen10,Gauthier13,Bordoloi14a,Nielsen15,Schroetter16,Nielsen18,Rubin18,Rubin18a}. These studies have found that \MgII\ absorbers trace the CGM of galaxies and possibly outflowing gas up to a distance of $\sim 100$ kpc \citep{Kacprzak08, Chen10}. This transition is therefore ideal to study the CGM of galaxies in groups and in isolation in the MUDF, for the first time in a very deep and complete dataset. 

The paper is structured as follows: first we present the MUDF survey strategy, the data reduction procedures, and the quality validation of the MUSE data (Section \ref{obs_datared}), and of the high-resolution UVES spectroscopy (Section \ref{obs_hires}). We then describe the procedures adopted to extract the sources and their properties (Section \ref{sec_galaxyprop}), and the reconstruction of the local environment by searching for groups of galaxies in the field (Section \ref{sec_groups}). In Section \ref{sec_mgiiabs}, we describe a novel method to fit the high resolution quasar spectra to extract metal absorption profiles and we present our results on the correlation of absorbers and galaxies in groups and in isolation. We conclude with a discussion of these results (Section \ref{sec_discussion}) and with a summary of our findings (Section \ref{sec_conclusions}). 

Throughout this paper, we assume a flat $\Lambda$CDM cosmology with $H_0 = 67.7~{\rm km~s^{-1}~Mpc^{-1}}$ and $\Omega_m = 0.307$ \citep{Planck16}. All magnitudes are expressed in the AB system, distances are in proper units, and we assume a \citet{Chabrier03} stellar initial mass function.

\section{MUSE Observations} \label{obs_datared}

\subsection{Survey strategy and current status}

\begin{figure}
    \centering
    \includegraphics[width=0.50\textwidth]{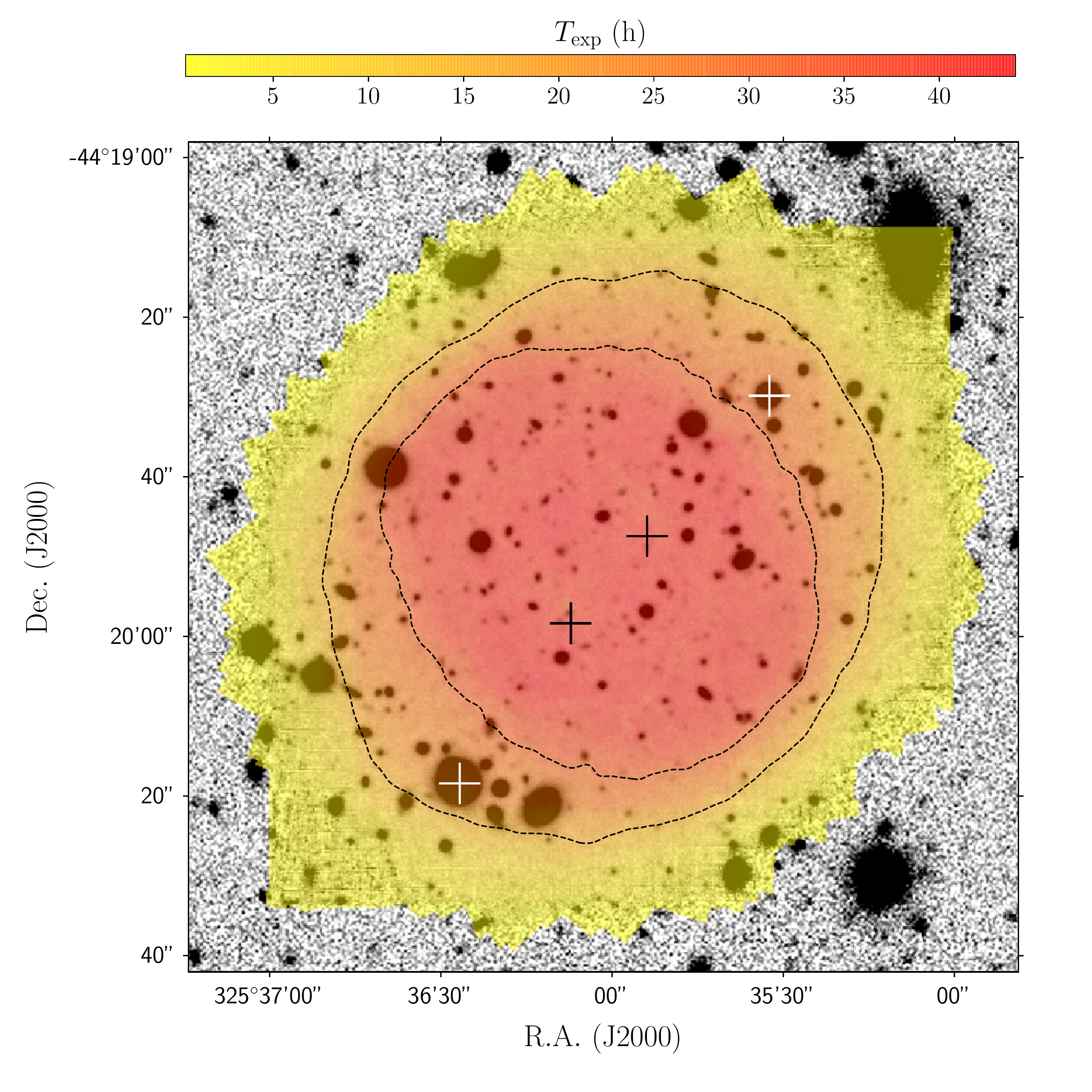}
    \caption{Exposure map (colour scale) of the MUSE observations in the MUDF overlaid on an optical white light image from the MUSE data where available or from the Dark Energy Survey { combined} $g,r,i$ images elsewhere { (shown in grey on the same surface brightness scale). North is up and East is left.} The outer and inner black dashed contours mark the regions with at least 15 hours and 30 hours of exposure, respectively in this partial dataset. { The black and the white crosses mark the pointing centres and the position of the two quasars, respectively}.}
    \label{fig:MUDFexpmap}
\end{figure}

The science goals of the MUDF programme include the study of the galaxy population around and along the line of sight to the quasar pair, a deep search for \lya\ emission from the putative filaments which are expected to connect the two quasars at $z\approx 3.22$.
%, and the study of strong absorbers at $z\approx 3$. 
Because the projected distance of these quasars is $\approx 62$ arcsec on the sky, a single Wide Field MUSE pointing of 60 arcsec on a side (the exact shape is trapezoidal) would not allow a full mapping of the area of interest. For these reasons, we designed an observational strategy that includes two heavily-overlapping pointings: { North-West and South-East (hereafter named simply North and South)}, the centres of which are shown as black crosses in Figure \ref{fig:MUDFexpmap}. Throughout the entire survey, we plan to collect $\sim 200$ frames dithered around each of these centers. The nominal exposure time of each frame is 1450s and different frames include small on-sky dithers ($\approx 3-4~$ arcsec), as well as 10 deg rotations of the instrument to reduce systematic errors arising from the different response of the 24 spectrographs and detectors of MUSE. While multiple exposures will be taken at the same rotation angle, the sequence of dithers has been designed to ensure that those exposures will not have the same centre and orientation. 

The final survey footprint will be ellipsoidal, with a $\approx -45$ deg position angle (North through East), and a major and minor axis of $\approx 110$ and 90 arcsec respectively. At the time of writing, we have reached $\sim 35\%$ completion, and Figure \ref{fig:MUDFexpmap} shows the exposure map generated with these data overlaid on a combination of white-light images from the MUSE data itself and from the Dark Energy Survey (DES) Data Release 1 \citep{Abbott18} outside the MUSE footprint. The survey design prioritises the collection of deeper data in the area between the two quasars, where our science goals warrant maximum sensitivity. The outer black dashed contours mark the regions with at least 15 hours and 30 hours of exposure, respectively. 

The instrument setup makes use of the MUSE Wide Field Mode with extended wavelength coverage in the blue (4650$-$9300$\AA$), in order to search for \lya\ emitting galaxies down to $z\approx 3$. We also take advantage of the Ground Layer Adaptive Optics module (GALACSI) which uses artificial laser guide stars to improve the image quality by partially correcting for atmospheric turbulence. In this way, the observations can be performed under a wider range of natural seeing conditions without compromising the image quality of the final mosaics. However, this setup implies that we cannot use data between 5760$\AA$ and 6010$\AA$, a range affected by the sodium line generated by the laser beam. Moreover, laser induced Raman scattering of molecules in the atmosphere creates emission lines outside this range. These lines are removed as part of the sky subtraction data reduction steps with no impact on the data quality.

In this paper, we present results obtained with the first 44.1 hours of observations. The data are collected as part of ESO programme 1100.A-0528. The first run (A) was executed in visitor mode on 16th August 2017, where we acquired 19 exposures of 1270s each in dark-time. The conditions were excellent with clear skies and an image quality of 0.4-0.6 arcsec in the final reconstructed frames. After this successful run, further dark-time observations have been executed in service mode (runs B and C) under similarly good observing conditions. As of June 2019, we have acquired 93 frames with an exposure time of 1450s each. In total we have acquired 159ks of data, corresponding to 44.1h. The image quality on the final coadd has a FWHM=0.57 arcsec as measured by Moffat profile fits to point sources.

%The top panel in Figure \ref{fig:MUDFpsfdistr} shows the distribution of the image quality measured by fitting Moffat profiles to multiple point sources in each frame.  The vertical red dashed line corresponds to FWHM=0.57 arcsec as it is measured on the final coadd of all the exposures.  In the bottom panel we show the average flux measured in these point sources in each frame relative to the flux in the final coadd.

%\begin{figure}
%    \centering
%    \includegraphics[width=0.45\textwidth]{MUDF_qc_hist_112.pdf}
%    \caption{Top panel: Distribution of the image quality in individual exposures, measured as the FWHM of Moffat profiles fit to multiple point sources in each frame.  The vertical red dashed line shows the FWHM measured on the final coadd of all the exposures. Bottom panel: distribution of the average flux measured on point-sources in each frame relative to the average flux for the same sources in the final coadd. }
%    \label{fig:MUDFpsfdistr}
%\end{figure}

\subsection{Data reduction}

The data reduction procedure follows the methodology described in previous works by our team \citep[][]{Fumagalli16,Fumagalli17,Lofthouse19}. In brief, we use the ESO pipeline \citep[v2.4.1][]{Weilbacher14} to reduce the calibrations (bias, flat, arcs, and standard stars) and apply them to the individual exposures. We also use this software to resample the detector values into cubes that are then sky subtracted using models of the sky continuum and sky lines that are matched to the data extracted from the darkest pixels in each frame. The individual exposures are then aligned by using the position of point sources in the field, and we generate a first coadd of the frames. We register this initial coadd to match the WCS coordinates of the two quasars from the Data Release 2 of the {\it GAIA} survey \citep{Gaia18}. The combination of relative offsets (arising from dithers) and this absolute WCS offset (arising form the pointing accuracy of MUSE) is then propagated into the pixel-tables of each exposure. We then reconstruct the cubes again for each exposure, this time on a pre-defined spatial grid of 540$\times$540 pixels, with each voxel (volumetric pixel) measuring 0.2 arcsec in the spatial direction and 1.25\AA\ in the spectral direction. The size of the grid has been chosen such that all the spatially aligned exposures of the programme will fall onto it, which eliminates the need for re-projecting the data when new observations are acquired.  

The cubes produced by the ESO pipeline are not yet fully science-grade, as they are still affected by an uneven spatial illumination and by residuals from sky lines subtraction \citep{Bacon17}. To correct for these effects, we post-process the individually aligned, reconstructed and un-skysubtracted exposures using routines from the {\sc CubExtractor} package (v1.8, {\sc CubEx} hereafter, Cantalupo in prep.; see \citealt{Cantalupo19} for a description of the algorithms). The adopted procedure follows earlier work using MUSE data \citep{Borisova16, Fumagalli16, Fumagalli17}. 
%The details of the algorithms adopted and their performances will be presented in a forthcoming paper (Cantalupo in prep.). 

In brief, we use the {\sc CubeFix} tool to correct residual differences in the relative response of the 24 MUSE IFUs and of individual slices, which are not fully corrected by flat-field calibration frames. {\sc CubeFix} improves the illumination uniformity by measuring the average illumination in each stack 
(the MUSE FoV is composed of 24 IFUs which are further made of 4 stacks of 12 slices), as a function of wavelength on white-light images generated on-the-fly from the cube. We then use the {\sc CubeSharp} tool for sky subtraction. The algorithm performs a local sky subtraction, including empirical corrections of the sky line spread function (LSF).  The combination of these two routines is applied twice, by using the first illumination-corrected and sky-subtracted cube to mask the sources in the field for the second iteration. The masking step is critical to measure the true instrumental illumination in each stack and to achieve a high-quality sky subtraction. 

After this first double pass of the {\sc CubEx} tools on the individual frames, we coadd them with mean statistics and 3$\sigma$ clipping to generate a deep white light cube. This cube is then used to detect and mask the sources and the deep mask is then fed into {\sc CubeFix} and {\sc CubeSharp} for a final run. After this step we measure the FWHM of point sources, their average flux and their position with respect to the {\it GAIA} astrometry using Moffat fits to the white light images of each exposure. The distribution of these values across all the exposures are then inspected to ensure that all the frames are correctly aligned and flux calibrated. The photometry is consistent at a 4\% r.m.s. level across frames and the astrometric precision is within 0.05 and 0.03 arcsec in R.A. and Dec. respectively, i.e. $<10\%$ of the spatial resolution. Moreover, no frame has been identified to be a $>3 \sigma$ outlier on any of these metrics and therefore we include all frames in the final combine. 

Lastly, we combine the individual cubes with a $3\sigma$ clipping rejection of outliers and both with mean and median statistics. We also generate mean combines obtained with two independent halves of the exposures. These products are useful to correctly identify weak emission-line sources in the cube from residual artefacts (e.g. residuals of cosmic ray hits), which are likely to appear only in one of the two independent combines. { The {\sc CubEx} reduction pipeline significantly improves the quality of the illumination uniformity across the entire observed area. We quantified the flatness of the illumination by comparing the standard deviation of the flux in sky pixels of the white-light image from the {\sc CubEx} processing relative to the ESO reduction, finding a ratio of 0.23. This means that the products used in this work are four times deeper than what would have been possible to achieve with the standard pipeline.}

\subsection{Noise characterisation}
\begin{figure}
    \centering
    \includegraphics[width=0.48\textwidth]{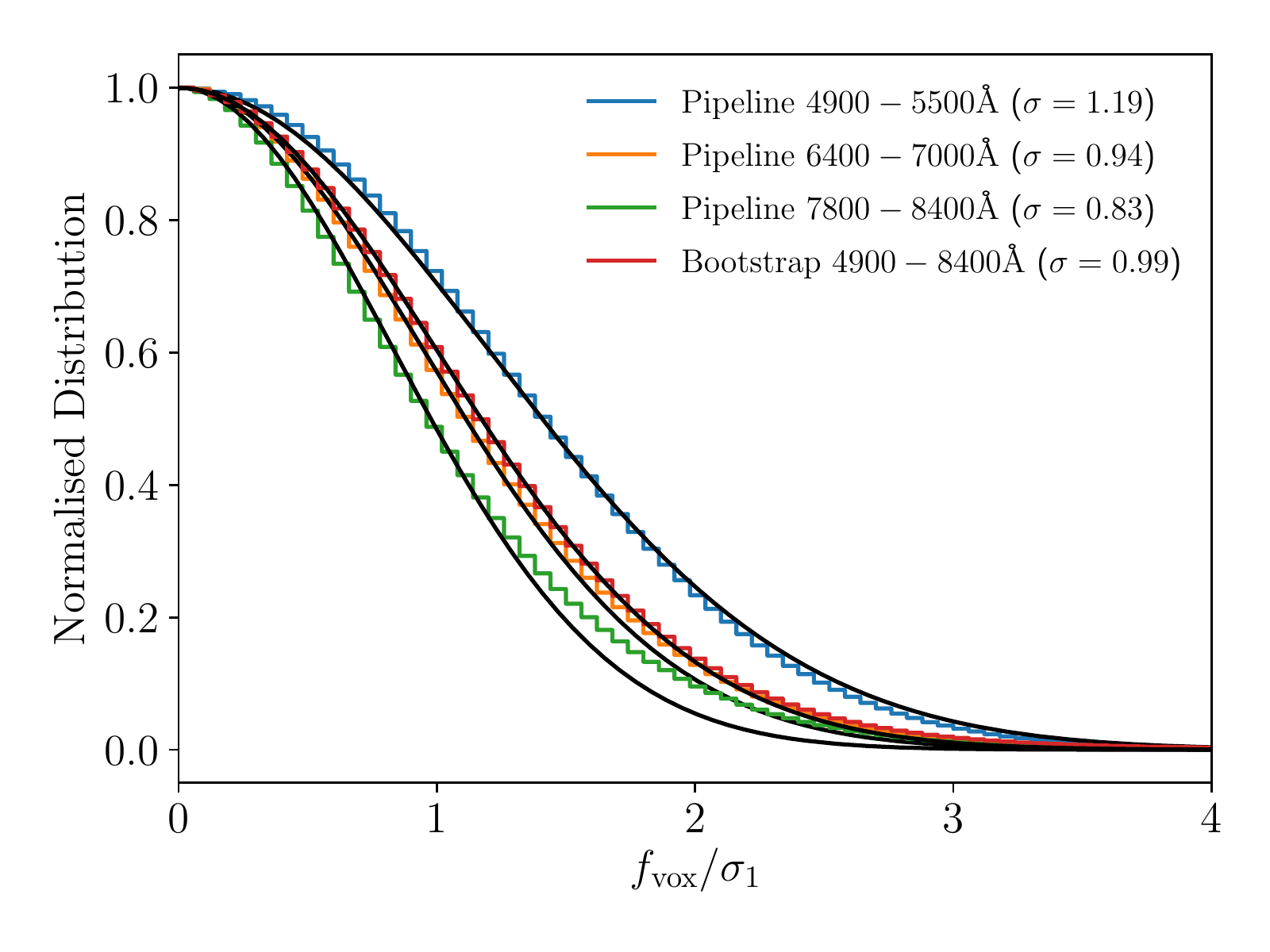}
    \caption{Histograms of flux in voxels values, normalised by the voxel standard deviation from the ESO+{\sc CubEx} pipeline in three wavelength ranges (blue, orange, and green), and in the full wavelength range after reconstructing the variance cube using a bootstrap resampling technique (red). Shown with black lines are Gaussian fits to these distributions, with the resulting standard deviation listed in the legend. Only spatial pixels free from continuum sources are included in this analysis. }
    \label{fig:rmshisto}
\end{figure}

During each step of the reduction process, the Poisson noise from detector counts is propagated and then combined into a cube that contains the variance of the resampled flux values. However, during several steps, including the drizzle interpolation onto the final grid, the propagated variance departs from accurately reproducing the effective standard deviation of individual voxels in the final data cube \citep[see][]{Lofthouse19}. In Fig.~\ref{fig:rmshisto} we show the flux distribution of voxels ($f_{\rm vox}$) normalised by the pipeline error ($\sigma_1$) in each pixel within three wavelength intervals that are increasingly affected by atmospheric sky lines (namely $4900-5500~\rm \AA$, $6400-7000~\rm \AA$, and $7800-8400~\rm \AA$, in blue, orange, and green, respectively). Once sources are masked, the distribution is expected to approximate a Gaussian (the black lines are Gaussian fits to the distributions), with standard deviation of unity. Instead,  Figure \ref{fig:rmshisto} shows that the pipeline error underestimates the true standard deviation in regions free from sky lines (blue histogram, with $\sigma=1.19$), while it overestimates it in regions more contaminated by skylines (green histogram, with $\sigma=0.83$). { Moreover, the distribution of $f_{\rm vox}/\sigma_1$ in the wavelength interval $7800-8400~\rm \AA$ shows the largest departure from a Gaussian distribution, an effect that could be attributed to the second-order contamination of the spectra due to our use of the MUSE extended wavelength mode.}

We overcome this issue by bootstrapping the combination of individual exposures for each pixel to accurately reconstruct the noise in the final mean, median, and half-exposure cubes. For this, we use 10,000 realisations of the bootstrap procedure to produce a variance cube. We then inspect the flux distribution of pixel values divided by the standard deviation of this new error, finding a distribution much closer to Gaussian, although still offset by a few percent from unity. We attribute this offset to a non-Gaussian distribution of the $f_{\rm vox}$ values. Due to this small offset, we further rescale the bootstrap variance cube to obtain a distribution of $f_{\rm vox}/\sigma_1$ that is unity on average, as shown by the red line in Figure \ref{fig:rmshisto}. We also note that by adopting this improved variance cube, the trend with wavelength which affects the propagated variance is also removed.

\begin{figure}
    \centering
    \includegraphics[width=0.50\textwidth]{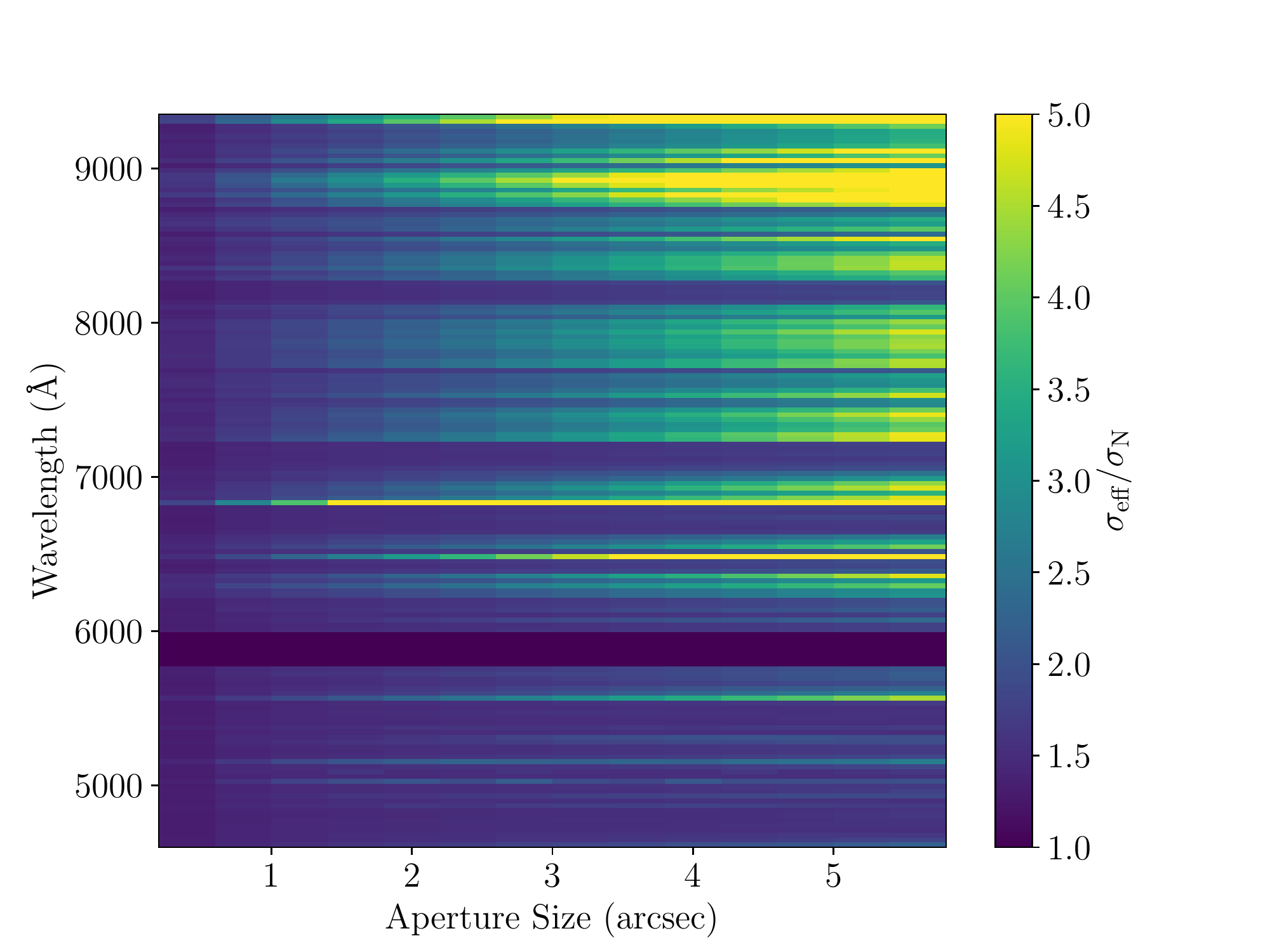}
    \caption{Ratio between the flux dispersion in apertures of varying size ($\sigma_{\rm eff}$) and the error derived propagating the variance ($\sigma_{\rm N}$), as computed across the MUDF datacube in a 5\AA\ window (4 pixels) as a function of wavelength and aperture size. The dark stripe just below 6000\AA\ is the region without valid data due to the scattered light from the AO laser. }
    \label{fig:covariance2dmap}
\end{figure}

Lastly, we derive a model for the correlated noise arising from the resampling of the pixel tables onto a final grid, as described in \citet{Lofthouse19}. This model represents the correction that needs to be applied to the propagated error for a source in a { square} aperture of $N$ spatial pixels on a side, $\sigma_{\rm N}$, to recover the effective noise, $\sigma_{\rm eff}$. In the spectral direction we use an aperture of 4 pixels ($\approx 5 \AA$) which is generally appropriate for narrow emission lines in galaxies. Figure \ref{fig:covariance2dmap} shows this correction in bins of wavelength and aperture size. While the dependence with aperture size is a smooth and monotonic function, the wavelength trend is not trivial to understand. It does not correlate to the brightness of sky lines, and we hypothesize it might be driven by the opto-mechanical design of the instrument, by our observing strategy, or more likely by a combination of the two. Regardless of the physical origin, when using these data to search for line emitters, we will use a second-order polynomial fit that describes $\sigma_{\rm eff}/\sigma_{\rm N}$ as a function of the aperture size, for the wavelength bin where each source is found. 

\section{High-resolution Quasar spectroscopy} \label{obs_hires}

As part of the MUDF survey, we are collecting a set of ancillary data to complement the MUSE observations. In this work, we make use of { UVES} high-resolution spectroscopy of the two quasars, J214225.78-442018.3 { (also known as Q2139$-$4434, and hereafter QSO-SE)} at $z = 3.221 \pm 0.004$ and J214222.17-441929.8 { (Q2139$-$4433, hereafter QSO-NW)} at $z = 3.229 \pm 0.003$ \citep{Lusso19}, for which we provide a description of the data acquisition and data reduction. 

Some UVES data for the two quasars already exists in the ESO archive \citep[PIDs 65.O-0299, 68.A-0216, 69.A-0204,][]{DOdorico02}. 
These data cover the wavelength range $\approx 4100-9000~$\AA, with a gap between $\approx 7400-7500~$\AA, { due to the gap between the two CCDs in the red arm of UVES}. The S/N of the bright quasar is $\approx 25$ per pixel across most of the wavelength range, while the fainter quasar has a S/N $\approx 8$ per pixel. To increase the wavelength coverage (down to $\approx 3600~$\AA\ and to fill the current gaps) and to increase the S/N of the fainter quasar, we have been awarded a total of 23 hours of new UVES observations (PID 102.A$-$0914). At the time of writing, observations for \qsoone\ have been completed for a total of 7h on-source, while for \qsotwo\ only 15.5h were obtained, and 17h are still to be observed. In this paper, we therefore make use of the full dataset for the brighter quasar, relying only on a partial dataset for the fainter one. We will describe the spectra obtained from the complete observations in a forthcoming paper.

%\begin{table}
%\caption{Journal of the UVES observations}
%\begin{minipage}{75mm}
%\label{tab:uves}
%\begin{tabular}{l l  c c c r}
%\hline  
%\multicolumn{6}{c}{\qsoone, $z_{\rm em} = 3.221$} \\
%Prop. ID & Date & Setup (nm) & Slit & No. &  T$_{\rm exp}$  \\ 
%\hline
%65.O$-$0299 & Sep 2000 & $474-750$ & 1.2 & 2 & 7200 \\
%69.A$-$0204 & Aug 2002 & 580 & 1.0 & 2 &  9000\\
%0102.A$-$194 & Oct 2018 & $390-564$ & 1.0 &3 & 8985\\
%\hline
%\multicolumn{6}{c}{\qsotwo, $z_{\rm em} = 3.229$} \\
%Prop. ID & Date & Setup (nm) & Slit & No. &  T$_{\rm exp}$ \\ 
%\hline
%65.O$-$0299 & Sep 2000 & $474-750$ & 1.2 & 2 & 9000 \\
%68.A$-$0216 & Oct 2001 & $470-750$ & 1.2 & 2 & 9000 \\
%69.A$-$0204 & Aug 2002 & 580 & 1.0 & 4 & 21600 \\
%0102.A$-$194 & Oct-Nov 2018 & $390-564$ & 1.0 & 3 & 16185 \\
%\hline
%\end{tabular}
%\end{minipage}
%%\end{center}
%\end{table}

All data were reduced with the current version of the UVES pipeline (v. 5.10.4), using default parameters and procedures.  At the end of the standard reduction process, the non-merged, non-rebinned spectra were reformatted with 
a custom script and input to the ESPRESSO Data Analysis Software \citep[DAS,][]{Cupani16} for the final operations of coaddition and continuum fitting. This step avoids multiple rebinning of the spectra, which would introduce correlations in the error array.
Spectra have been normalized to the continuum estimated by the ESPRESSO DAS. This software fits a cubic spline to the spectrum redward of the \lya\ emission. In the \lya\ forest, the fit is iteratively improved by the simultaneous fit of the \lya\ absorption lines (for more details see \citealt{Cupani16}).  
The final continuum-normalised spectra were then rebinned to a constant velocity step of 2.5 \kms. 

\section{The galaxy population in the MUDF} \label{sec_galaxyprop}
In this Section we characterize the galaxy population detected in the MUDF and describe the procedures used for the identification of continuum sources, the extraction of their spectroscopic redshifts, and the derivation of their physical properties through stellar population syntesis fits. 

\subsection{Source detection}
We identify continuum sources using the {\sc SExtractor} \citep{Bertin96} software on the white light image reconstructed from the MUSE datacubes. We input a variance image and we use a conservative threshold of 3$\sigma$ above the local { noise}, and a minimum area of 10 pixels for detection. The minimum deblending parameter {\sc DEBLEND\_CONT} is set to 0.0001, chosen to enable the detection of sources in crowded regions of the mosaic. We restrict source extraction to the area where we collected more than 10 exposures, corresponding to an observing time of $\approx 4$ hours, to avoid spurious detections at the noisy edges of the field of view. In future publications we will use deep {\it HST} imaging for source detection in the field. 

This procedure, identified 250 sources. For each of them, we extract the magnitude in the detection image ($m_{\rm MUSE}$) in an elliptical aperture with size equal to 2.5 times the \citet{Kron80} size from {\sc SExtractor}. We also reconstruct a 1D spectrum summing the spectra from pixels within the Kron aperture, also transforming the wavelength to vacuum. In both these procedures, we mask nearby sources whose segmentation map falls in the extraction mask to minimize the effects of blending. 

\begin{figure*}
    \centering
    \includegraphics[width=0.98\textwidth]{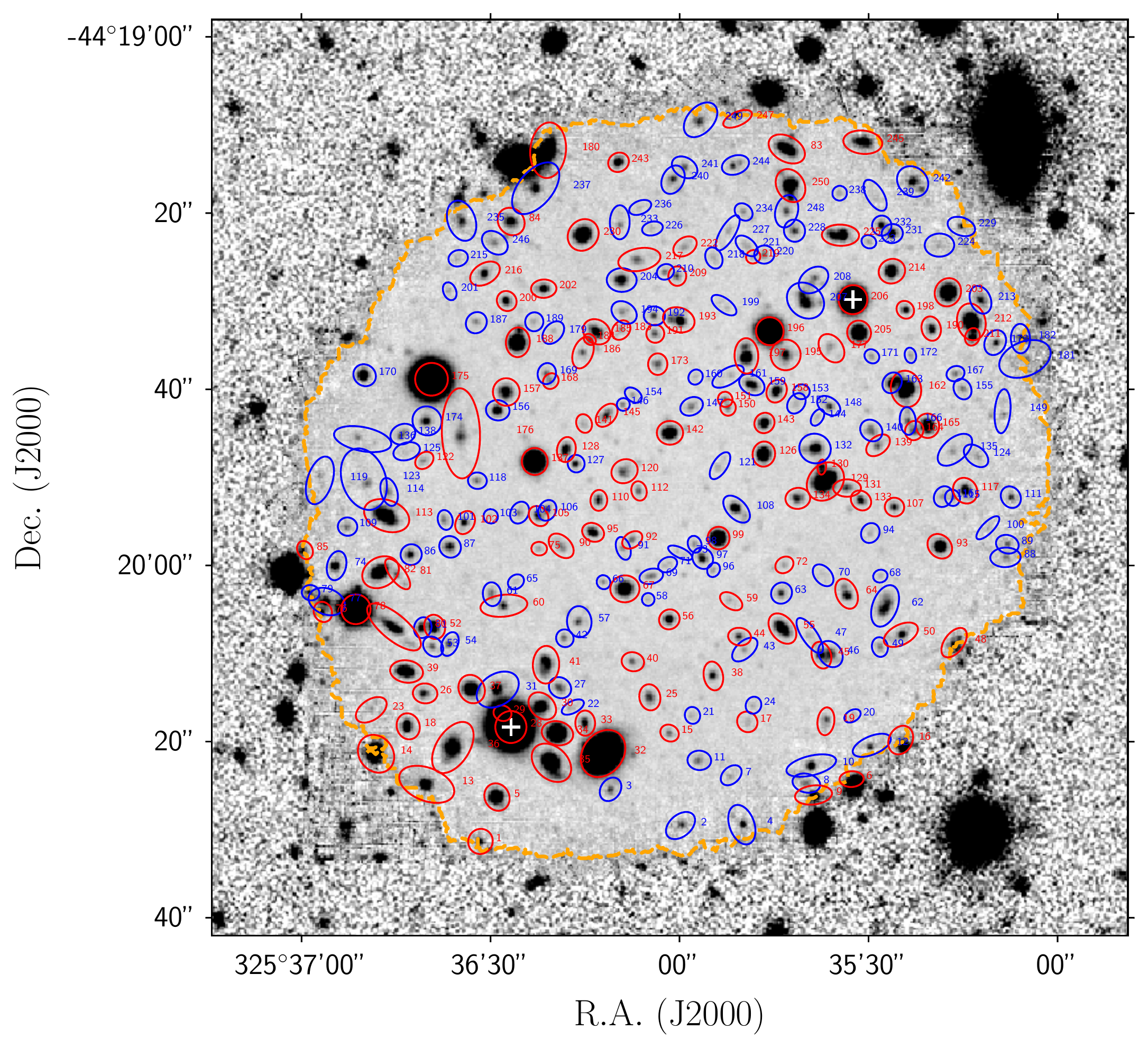}
    \caption{Reconstructed optical white-light image from the MUSE data (where available) or from the DES $g,r,i$ images (outside the MUSE FoV). { North is up and East is left}. The two quasars are marked with white crosses. Continuum-detected sources identified within the MUSE field are marked with apertures (red for sources with spectroscopic redshifts with confidence classes 4 and 3) with size equal to 6 times the Kron radius. The dotted orange contour encloses the region where the exposure time is $\ge 4$h.}
    \label{fig:musefov_sex}
\end{figure*}

We measure redshifts using the {\sc Marz} software \citep{Hinton16}, which we customize\footnote{This version is available at \url{https://matteofox.github.io/Marz}} with the inclusion of high-resolution synthetic templates for passive and star-forming galaxies at $z<2$ derived from \citet{Bruzual03} stellar population synthesis models. Following automatic template fitting, individual sources are inspected and classified by two authors (MFo and EKL) in four classes (4, secure redshift with multiple absorption or emission features; 3, good redshift with single but unambiguous feature; 2, possible redshift, based on a single feature; 1, unknown redshift). Typical redshift uncertanties are $\delta z \approx 0.0002 \times (1+z)$, corresponding to $\delta v \approx 60$ \kms. 

\begin{figure}
    \centering
    \includegraphics[scale=0.44]{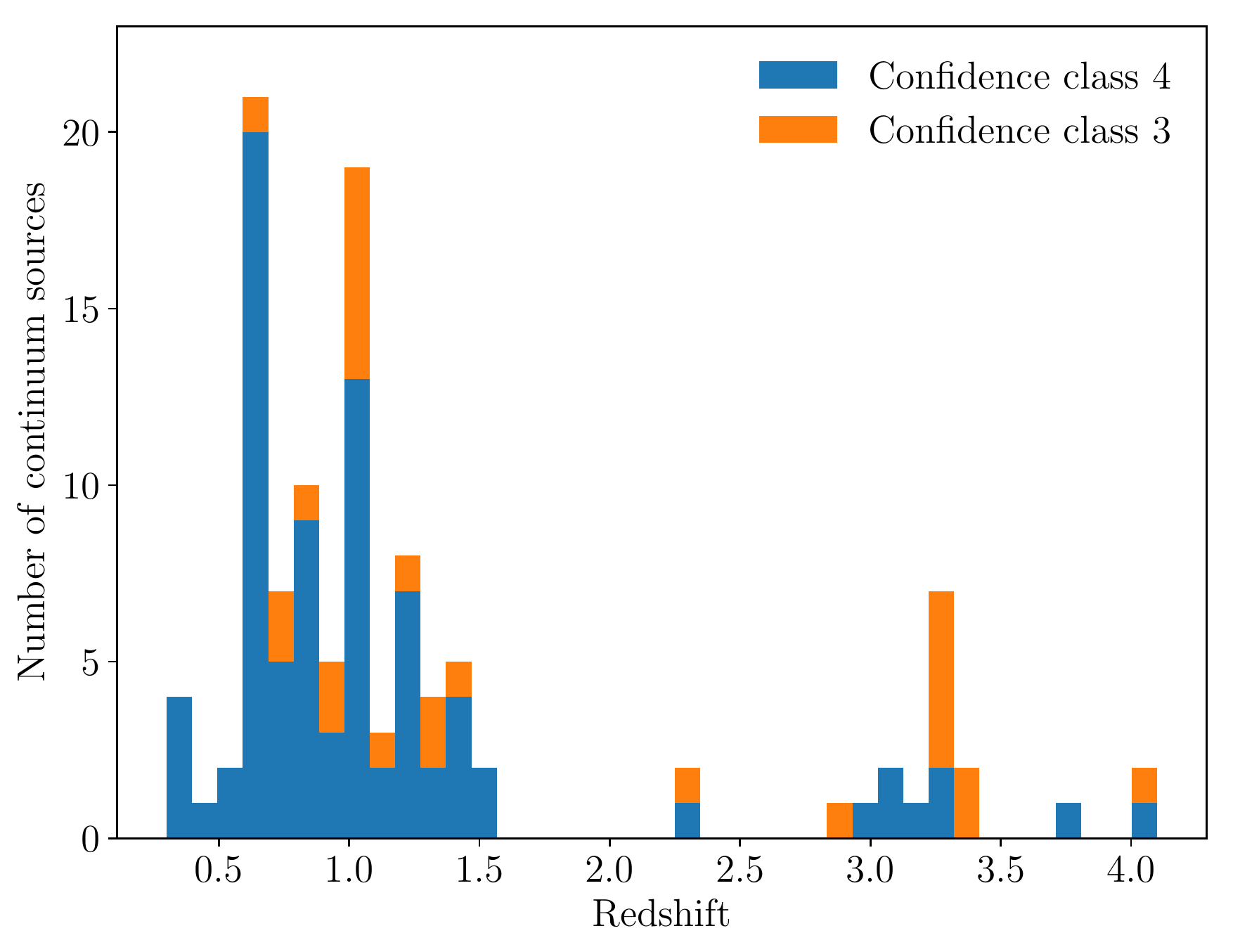}
    \caption{Distribution of spectroscopic redshifts with confidence classes 4 and 3 (see text for details) for continuum sources in the MUDF. }
    \label{fig:zdistr}
\end{figure}

The final redshift classification is presented in Table~\ref{tab:sourcesample}. Figure \ref{fig:musefov_sex} shows the position within the MUDF footprint of the continuum-detected sources and their {\sc SExtractor} IDs. Hereafter we make use only of the objects with a reliable spectroscopic redshift (classes 3 and 4) which are marked with red apertures. The redshift distribution of these { 117 sources (47\% of the detected sample)} is shown in Figure \ref{fig:zdistr}. The range $1.5<z<2.5$, known as the ``redshift desert'', does not have significant detections due to the absence of strong emission lines that would fall within MUSE coverage.
Only two galaxies are identified there thanks to strong \AlIII\ absorption lines in their spectra. This range, however, will be filled in by the ultra-deep near-infrared observations which we are collecting with {\it HST}/WFC3.

\begin{table*}
\centering
\begin{tabular}{ccccccccc}
\hline
ID & Name  & R.A. (J2000) & Dec. (J2000)& $m_{\rm MUSE}$ & $\sigma_{m_{\rm MUSE}}$ & Redshift & Confidence & $\log(M_*/{\rm M_\odot})$ \\
\hline
1 & MUDFJ214226.11-442031.3 & 325.60878 &$-$44.34204 &24.3 & 0.2 & 4.0441 & 4 & -\\ 
2 & MUDFJ214223.99-442029.5 & 325.59996 &$-$44.34154 &26.0 & 0.3 & -      & 1 & -\\ 
3 & MUDFJ214224.73-442025.4 & 325.60304 &$-$44.34039 &25.7 & 0.3 & -      & 1 & -\\ 
4 & MUDFJ214223.34-442029.4 & 325.59725 &$-$44.34151 &24.3 & 0.1 & -      & 1 & -\\ 
5 & MUDFJ214225.93-442026.3 & 325.60805 &$-$44.34064 &23.6 & 0.1 & 1.0530 & 4 & 10.24\\ 
\hline
\end{tabular}
\caption{The first five continuum sources in the MUDF extracted by {\sc{Sextractor}} with $S/N>3$. Column 1 shows the source ID, column 2 shows the source name. Columns 3 and 4 list the right ascension and declination of the sources followed by the MUSE white-light magnitude of the source in column 5 with its associated error (column 6). The redshifts obtained using {\sc{MARZ}} are shown in column 7 followed by their confidence (column 8). A confidence flag of 3 or 4 indicates reliable redshifts while flags 1 or 2 are for unknown or highly uncertain redshifts, respectively. Column 9 shows the stellar mass from stellar population fitting for sources with $z<1.5$. The full table is included as online only material.}
\label{tab:sourcesample}
\end{table*}

\subsection{Completeness}

To assess the completeness of our source extraction procedure, we inject mock sources of known magnitude into the detection { (white-light)} image and we assess their detection rate. 
First, we inject two dimensional Gaussian templates with a FWHM of 0.6~arcsec to simulate unresolved point sources at the resolution limit of the final MUSE mosaic. Then, we repeat the experiment using circular exponential profiles which are more appropriate for real disk-like galaxies. In this case we use an exponential scale length of 0.26~arcsec which corresponds to an effective radius of 5~kpc at $z\sim1$, which is typical for star-forming galaxies \citep{van-der-Wel14}. The intrinsic exponential profiles are convolved with a Gaussian kernel with a FWHM of 0.6~arcsec to account for the effects of the observational PSF, { and do not include the effects of the disk inclination}. In each iteration, we inject 80 mock sources (to avoid confusion and blending issues) in blank background regions and we repeat the detection procedure 10,000 times.

\begin{figure}
    \centering
    \includegraphics[scale=0.45]{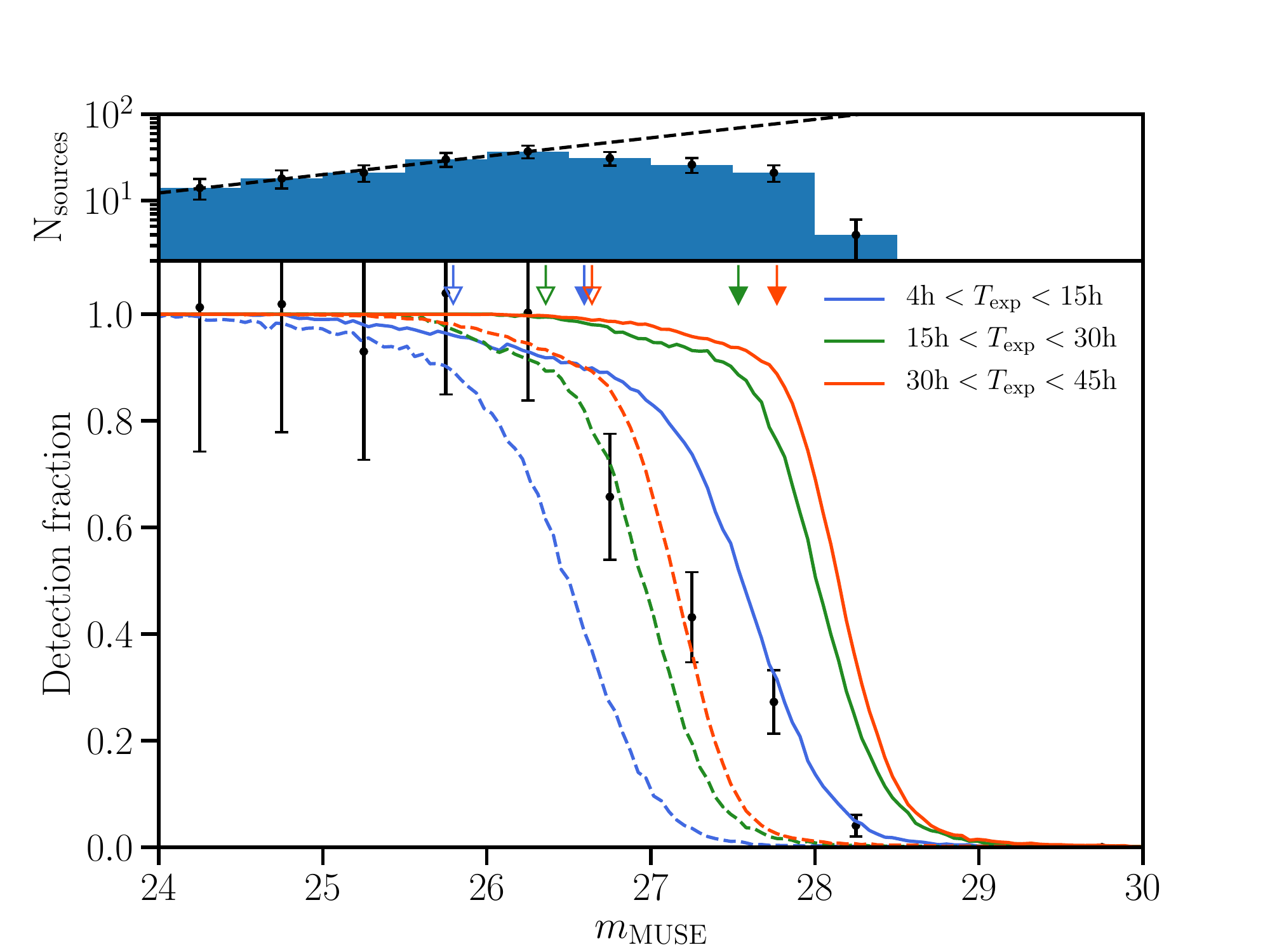}
    %\caption{Fraction of mock sources detected by {\sc Sextractor}. The solid lines are for the injection of Gaussian sources with a FWHM of 0.6~arcsec. The dashed lines are for sources with an exponential profile with scale-length equal to 0.26~arcsec convolved with a two-dimensional kernel to account for the effective PSF. Lines of different colours are for different exposure times in the MUDF map. The arrows (filled for point sources and empty for extended ones) mark the faintest magnitude at which we are 90\% complete in each bin of exposure time. }
    \caption{{ Top Panel: Number of galaxies extracted as a function of the magnitude in the detection image. The grey dashed line shows the expected number of galaxies determined from a linear fit to the region between $m_{\rm{MUSE}}>24$~mag and $m_{\rm{MUSE}}<26.5$~mag.}
    Bottom Panel: Fraction of mock sources detected by {\sc Sextractor}. The solid lines are for the injection of Gaussian sources with a FWHM of 0.6~arcsec. The dashed lines are for sources with an exponential profile with scale-length equal to 0.26~arcsec and PSF convolved. Lines of different colours are for different exposure times in the MUDF map. The arrows (filled for point sources and empty for extended ones) mark the faintest magnitude at which we are 90\% complete in each bin of exposure time. { The black points (with errorbars) show the empirical detection fraction determined as the ratio of the detected galaxies relative to the expected number from the linear fit in the top panel. }}
    \label{fig:simdepth}
\end{figure}

{ The bottom panel of} Figure \ref{fig:simdepth} shows the fraction of detected mock sources as a function of magnitude for sources injected at locations of the image with different exposure times. For point sources (solid lines) we reach fainter limits than for extended sources (dashed lines), due to their compactness. The arrows mark the faintest magnitude at which we are 90\% complete in each bin of exposure time. In the deepest bin, where we have so far collected between 30-45 hours of data, we reach $m_{\rm MUSE} \simeq 27.8~\rm mag$ and 26.6 mag for point-like and extended sources, respectively. { The top panel of Figure \ref{fig:simdepth} shows the number of detected galaxies as a function of the white-light magnitude, $m_{\rm MUSE}$. We fit a linear relation (dashed line) to the logarithm of the number counts in the region $ 24~\rm{mag} < m_{\rm MUSE} < 26.5~\rm{mag}$, where our survey is complete. We then show in the bottom panel the empirical detection fraction (black points) determined as the ratio of the detected galaxies relative to the expected number from the linear fit. Despite the low number statistics, we find that this empirical detection fraction is in between the limits defined by the two
mock experiments (point sources and extended sources), complementing and corroborating the approach based on mocks. 

Lastly we note that, to reach maximum depth, the detection image is obtained from the full MUSE wavelength range, which does not correspond to the most commonly used broad-band filters. To facilitate the comparison to other imaging surveys we have computed the following color corrections for a star-forming (passive) galaxy template at $z=1$: $m_{\rm MUSE}-r_{\rm SDSS} = -0.31 (-0.44)$, and $m_{\rm MUSE}-i_{\rm SDSS} = 0.04 (0.05)$.}

\subsection{Source photometry and ancillary data}
Taking advantage of the wide wavelength coverage of MUSE, we extract source photometry in four pseudo-filters with the goal of characterising the physical properties of the galaxy population. The width and the number of filters have been carefully selected to maximise our sensitivity to strong breaks in the galaxy spectra, while keeping a good S/N ratio in the measurements, and avoiding the gap in the spectra due to the AO laser filter. We convolve the MUSE datacube with the top-hat filters, whose ranges are given in Table \ref{tab:filterspec}, to generate an image and the associated variance. We then run a forced photometry algorithm using { apertures with radius 2.5$\times r_{\rm Kron}$} as defined above for the detection image.

\begin{table}
    \centering
    \begin{tabular}{lcc}
        \hline
        Filter Name & $\lambda_{\rm min}$ (\AA) & $\lambda_{\rm max}$   (\AA)  \\
        \hline
         MUSE Blue  & 4750 & 5700 \\
         MUSE Green & 6100 & 7100 \\
         MUSE Red   & 7100 & 8200 \\
         MUSE NIR   & 8200 & 9300 \\
    \end{tabular}
    \caption{Name and wavelength range of the top-hat filters defined for the photometric extraction.}
    \label{tab:filterspec}
\end{table}

The MUSE data probe the rest-frame near-UV to optical region of the galaxy spectra for sources at $z \lesssim 1.5$. While providing a good sensitivity of the recent star-formation activity and stellar content of the galaxies, these blue wavelengths are affected by dust extinction which could bias our reconstruction of the galaxy parameters. 

%\begin{figure*}
%    \centering
%    \includegraphics[width=0.98\textwidth]{id35_mudf_summary.pdf}
%    \caption{Example of the stellar population fitting results for source ID 35 in the MUDF. Top Panel: rest-frame MUSE spectrum (black line) and best fit model (red line). The fit residuals { (Data-Model)} are given below the main panel, and the shaded area shows the noise array. Bottom Left Panel: Spectral energy distributions extracted from the MC-SPF posterior (red lines) with the observed photometry overplotted in black.  The blue lines show the contribution of young stars (Age $< 10$~Myr) to the total template. Bottom Central panel: star-formation histories sampled from the MC-SPF posterior. Bottom Right Panel: marginalised likelihood maps for the Age and $\tau$ parameters of the SFH. The red lines show the median value for each parameter, while the black contours show the { 1$\sigma$ and 2$\sigma$} confidence intervals.}
%    \label{fig:sps_example}
%\end{figure*}

\begin{figure*}
    \centering
    \includegraphics[width=0.98\textwidth]{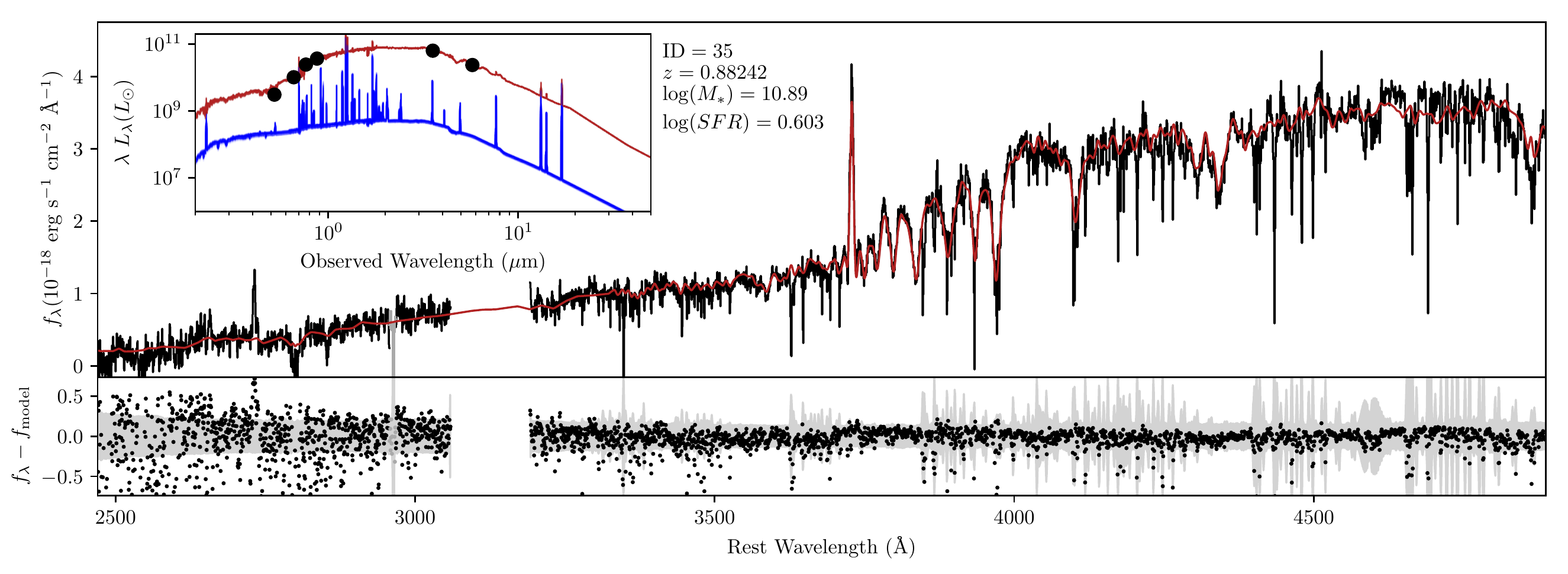}
    \caption{Example of the stellar population fitting results for source ID 35 in the MUDF. Rest-frame MUSE spectrum (black line) and best fit model (red line). The fit residuals { (Data-Model)} are given below the main panel, and the shaded area shows the noise array. The inset shows the spectral energy distributions extracted from the MC-SPF posterior (red lines) with the observed photometry overplotted in black.  The blue lines show the contribution of young stars (Age $< 10$~Myr) to the total template.}
    \label{fig:sps_example}
\end{figure*}

To mitigate this issue, we include in our analysis data taken with the IRAC \citep{Fazio04} instrument onboard the {\it Spitzer} \citep{Werner04} Space Telescope. IRAC imaging of an area of the sky including the MUDF has been taken during the cryogenic mission \citep[PID: GO-3699,][]{Colbert11} using the four channels at 3.6, 4.5, 5.8, and 8.0$\mu$m. However, due to the field of view offset of the IRAC camera, only the channels at 3.6, and 5.8$\mu$m cover the MUDF in its full extent, and we use them in this work. The total integration time per pixel of these observations are 1800s, and image FWHM is $\approx 1.8$~arcsec. 

To extract IRAC photometry from these images we use the {\sc t-phot} software \citep{Merlin15}, which optimally deals with the lower resolution of IRAC data in crowded fields by means of source priors from higher resolution imaging. We use the MUSE white light image, PSF model and source catalogue as the high resolution dataset. We then derive a Gaussian convolution kernel for each IRAC channel, to transform the MUSE PSF into the IRAC one. Lastly we resample the IRAC data onto the MUSE pixel grid and run {\sc t-phot}. Due to the shallowness of the IRAC data, only the relatively bright sources can be detected robustly, especially in crowded areas of the MUDF. In the end, we visually inspected the {\sc t-phot} results for each source to assess if results: i) can be used and the detection is at least with 3$\sigma$ significance; ii) can be used as an upper limit; iii) cannot be used because the source is faint and it is highly blended with a nearby bright source. In the first two cases (95\% of the sources we fit), we include the IRAC data in the estimates of the stellar populations as described in the following section.

\subsection{Stellar population parameters}
We characterize the physical properties of the galaxy population in the MUDF by jointly fitting the MUSE spectra and the photometric data derived from MUSE and IRAC as described above. These combined data sets allow us to derive reliable estimates for the star-formation history, stellar mass, current star-formation rate, and dust content of the galaxies. This multi-wavelength approach is instrumental in breaking the degeneracy that often arise between these parameters, most notably between stellar mass and dust extinction, or star-formation rate and galaxy age \citep{Wuyts11a, Fossati18}. In this work, we primarily use the derived stellar masses, and we postpone the discussion of results involving the other quantities to a follow-up paper, where we will re-assess the quality of these measurements by adding into the fitting procedure the  near-UV and near-IR data that will be collected with {\it HST} as part of programmes PID 15637 and 15968.

To infer the source properties, we fit the observed data to synthetic models using the Monte Carlo Spectro-Photometric Fitter (MC-SPF). A more complete description of the code is given in \citet{Fossati18}; here we briefly describe the procedure and the model set. First, we build a grid of synthetic stellar spectra from the \citet{Bruzual03} high-resolution models at solar metallicity. Following similar work that studied the properties of galaxies in deep fields \citep{Wuyts11a, Momcheva16}, we use exponentially-declining star-formation histories with $e$-folding times varying between 300 Myr and 15 Gyr, and galaxy ages varying between 50 Myr and the age of the Universe at the redshift of each galaxy. With the imposed minimum $e$-folding time, \citet{Wuyts11a} found that the SFR derived with this spectral energy distribution technique matches the one obtained from UV+far-IR photometry for systems { with low-to-intermediate star-formation rates ($SFR \lesssim 100~{\rm M_\odot~yr^{-1}}$), which are likely
to dominate the sources in our small field.} 

We further add nebular emission lines to the stellar template, by using line ratios from \citet{Byler18} scaled to the line luminosity using the number of Lyman continuum photons from the stellar spectra. The SFH and nebular emission grids are interpolated on-the-fly during the likelihood sampling. We assume a double \citet{calzetti00} attenuation law to include the effects of dust attenuation. Stars older than 10 Myr and emission lines arising from them are attenuated by a free parameter, $A_{\rm old}$, while younger stars are attenuated by $A_{\rm young} = 2.27\times A_{\rm old}$ to include the extra extinction occurring within the star forming regions.  

We jointly fit the photometric data points and the MUSE spectra at their native resolution. A third degree multiplicative polynomial is used to normalize the spectra to the photometric data and to remove large-scale shape differences between the models and the spectra, especially at the edges of the wavelength range where the MUSE response function is more uncertain. The multidimensional likelihood space is sampled by using {\sc PyMultiNest} \citep{Buchner14}, a python wrapper for the {\sc MultiNest} code \citep{Feroz08,Feroz13}. Figure \ref{fig:sps_example} shows an example of the results obtained with this fitting procedure for a galaxy in the MUDF. With the current wavelength coverage, some degeneracy remains between dust extinction and star-formation rate. Rest-UV data from HST will break this degeneracy, however, in this work we primarily make use of the stellar mass estimates from the fits which are found to be well converged and free from degeneracies with other  parameters.  

\section{Group Identification} \label{sec_groups}
The detection of a large number of sources with accurate spectroscopic redshifts in narrow redshift bins (see Figure~\ref{fig:zdistr}) is highly suggestive of the presence of overdense structures in the MUDF footprint. Indeed, over-densities spanning from compact groups to clusters and super-clusters of galaxies have been detected in all the fields which have been targeted by deep and extensive spectroscopic campaigns \citep[e.g.][]{Yang07, Scoville07, Kovac10, Diener13, Balogh14, Fossati17, Galametz18}. We therefore proceed to systematically identify galaxy groups in the MUDF. 

There is a rich collection of literature on finding groups in spectroscopic redshift surveys, with most methods based on a Friends-Of-Friends approach \citep{Huchra82, Berlind06, Knobel09, Knobel12, Diener13}. These methods link galaxies into structures by finding all objects connected within linking lengths $\Delta r$ (a physical distance) and $\Delta v$ in redshift space. The choice of these parameters is driven by the need to balance the competing  requirements of identifying all group members without over-merging different groups, avoiding interlopers, and taking into account redshift uncertainties. 
In this work we search for galaxy groups at $0.5<z<1.5$, where the lower limit is dictated by the small volume probed at lower redshift and the upper limit is given by the redshift desert. In this range, we have highly accurate spectroscopic redshifts for all the galaxies that we aim to connect into structures. We use $\Delta r = 400$ kpc and $\Delta v = 400$ \kms, following \citet{Knobel09} in the same redshift range. Similar to previous works, we define a galaxy group to be an association of three or more galaxies \citep{Knobel12, Diener13}, { independently of stellar mass or observed magnitude}. 

\begin{table*}
    \centering
    \begin{tabular}{ccccccccc}
    \hline
ID & <R.A.>  	& <Dec.>  & <z>     & $\rm{N_{gal}}$ &   $\log({M_{\rm halo}}/{\rm{M_\odot}})$ & $r_{\rm vir}$ & $W_{2796}$ & $W_{2796}$\\
 & J2000 & J2000 & & & & kpc & (QSO-SE) $\AA$ &(QSO-NW) $\AA$\\
\hline
1  & 325.602433 & -44.332283 & 0.67837 & 11 	  &   12.0			   & 160	& $0.85^{+0.03}_{-0.04}$ & $<0.08$	 \\ %4
2  & 325.598267 & -44.332865 & 0.68531 & 5  	  &   11.2 			   &  87    &  $<0.02$  & $<0.07$ \\ %5
3  & 325.593548 & -44.330041 & 0.78491 & 3  	  &   10.8 			   &  61	&  $<0.02$	& $<0.06$ \\ %6
4  & 325.597824 & -44.330187 & 0.88205 & 6  	  &   12.9 			   & 293	& $1.34^{+0.02}_{-0.01}$ &$0.43^{+0.01}_{-0.01}$	 \\ %9
5  & 325.604431 & -44.331396 & 1.05259 & 15 	  &   13.4 			   & 419	& $1.67^{+0.02}_{-0.02}$ & $<0.06$	 \\ %11
6  & 325.605063 & -44.331111 & 1.15524 & 3  	  &   12.1 			   & 138	& $1.16^{+0.06}_{-0.13}$ & $<0.06$	 \\ %12
7  & 325.602279 & -44.328677 & 1.22849 & 4  	  &   11.6 			   &  96	& $<0.02$ &  $<0.06$ \\ %14
    \end{tabular}
    \caption{Properties of the groups identified in the MUDF. The table lists: the group ID; the average R.A., Dec. and redshift of the group members; the number of members in each group; the halo mass derived from the stellar-to-halo mass relation of \citet{Moster10}; the virial radius (R200) for that halo mass; the equivalent width of the \MgII\ $\lambda$2796\AA\ absorption line associated with the group (or 2$\sigma$ upper limit) { along the \qsoone\, and \qsotwo\ sightlines.}}
    \label{tab:groups}
\end{table*}

Following this procedure, we find seven groups in the MUDF footprint, and their properties are listed in Table~\ref{tab:groups}. We estimate the group halo mass by summing the stellar mass of the group galaxies \citep[see][for a validation of the method]{Yang07} and using the stellar to halo mass relation from \citet{Moster10} at the redshift of the group. We also report in the table the geometric centre of the group, the average redshift of its members { (not weighted by other galaxy parameters)}, and the virial radius calculated from the halo mass. 
Assuming a typical uncertainty on the total stellar mass of 0.15 dex \citep{Conroy09, Gallazzi09, Mendel14}, this turns into an error on the virial mass of 0.20-0.25 dex depending on the local gradient of the stellar- to halo-mass relation.  This uncertainty corresponds to an error of $\sim 50-60$ kpc on the virial radius. The virial radius is further affected by the implicit assumption of virialization of the group halo. With only a few members per group, this can not be guaranteed as we might be observing groups in formation. Given these caveats, the estimated virial radii should only be taken as indicative values, especially for structures with less than 10 members. 

\begin{figure*}
    \centering
    \includegraphics[width=0.98\textwidth]{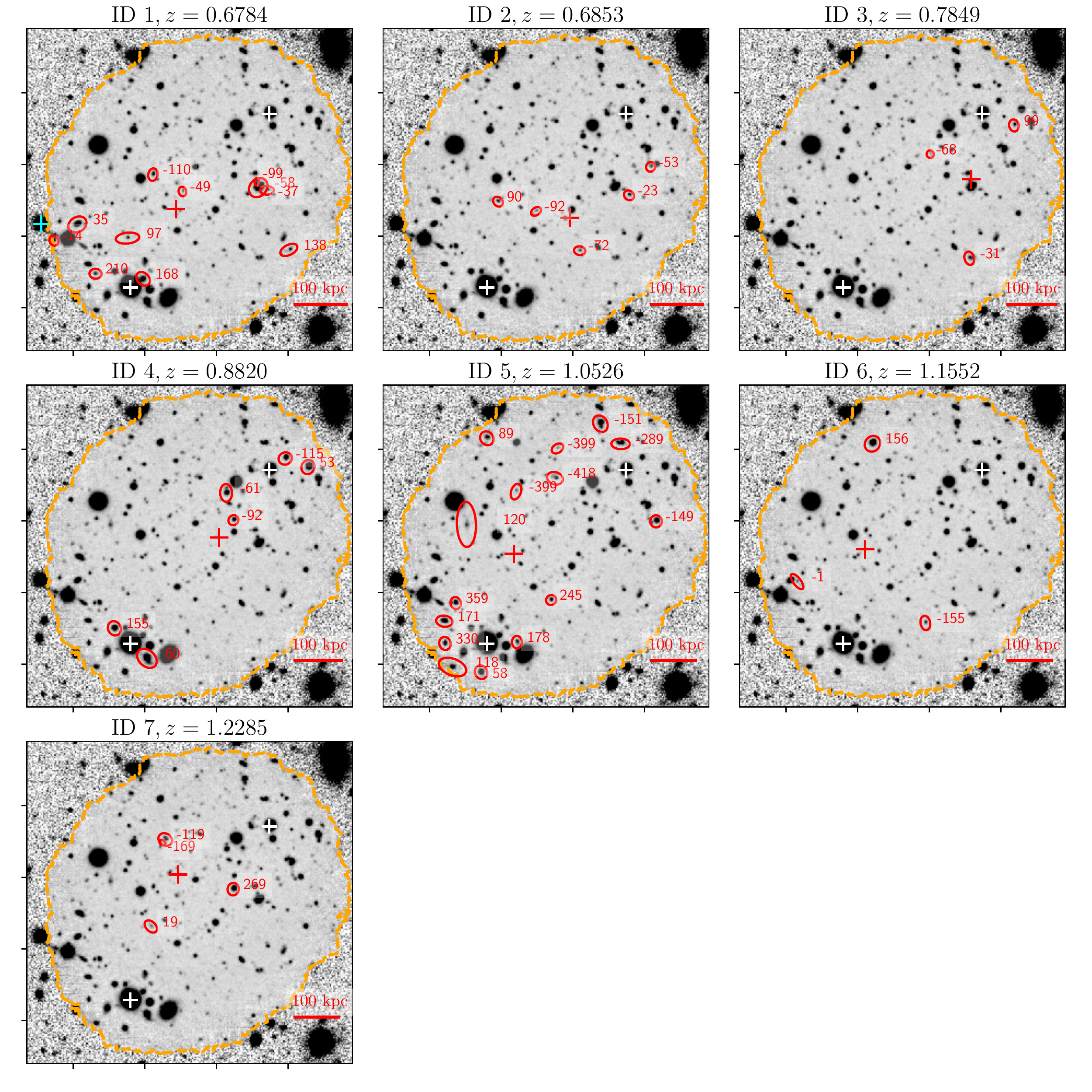}
    \caption{Gallery of the location within the MUDF of the galaxies in groups, marked with ellipses as in Figure \ref{fig:musefov_sex}. The white crosses mark the position of the background quasars, while the red crosses in each panel indicate the geometrical centre of the group. The cyan cross in the top left panel marks the position of a serendipitously detected quasar that shows absorption at the redshift of this group (see Section \ref{sec_connectingprop}). The number next to each galaxy shows the velocity offset with respect to the redshift of the group. The red segment in the bottom right corner of each panel marks a 100 kpc scale at the redshift of each group.}
    \label{fig:groups_gallery}
\end{figure*}

At the redshift of the groups, we search for strong line emitters which are too faint in continuum to be detected. We run {\sc CubEx} on the continuum-subtracted cube and we use the detection parameters defined in \citet{Lofthouse19}, i.e. voxel $S/N > 3$, minimum number of voxels of 27 and minimum number of channels in wavelength of 3. We found no line emitter that is not associated with a detected continuum source. However it remains possible that deeper exposures would lead to a secure detection of more redshifts from continuum features, possibly increasing the number of groups or including fainter members. 
Figure \ref{fig:groups_gallery} shows the location of the detected galaxies in groups within the MUDF footprint.

\subsection{Going beyond the survey edges}
The unprecedented depth of the MUSE observations in the MUDF comes at the price of a relatively small survey area. Therefore, it is possible that at least some of the groups identified in the previous section are part of a larger scale structure, or that some of the group members are missed by our footprint. Unfortunately, this area of the sky is not covered by wide-area archival spectroscopy, thus we can only attempt to characterise the large scale environment around the MUDF with photometric redshifts (photo-z). 

For this purpose, we downloaded source and magnitude catalogues from the DES DR1 \citep{Abbott18}, and we computed photo-zs from the $g,r,i,z$ fluxes using the {\sc EAZY} code \citep{Brammer08}. Following \citet{Hoyle18} we do not include $Y-$band magnitudes, because the observations are too shallow to improve the photo-z estimate. These authors also studied the photo-z uncertainty on a smaller region of the DES footprint, finding that they range from $\sigma_z = 0.1$ at $z=0.5$ to $\sigma_z = 0.4$ at $z=1.5$. For each group identified in the MUDF, we compute the galaxy density in an aperture with radius 1 Mpc centred on the geometrical centre of the group. The density is defined as the sum over all the galaxies within this aperture of the fraction of the DES photo-z probability density function that falls within $\Delta v = \pm 5000$\kms\ of the redshift of the group. This approach optimally takes into account the variable photo-z accuracy as a function of redshift and galaxy magnitude \citep{Kovac10, Cucciati14}. We then define two control apertures of the same size and velocity depth bracketing in redshift the previous aperture, but not overlapping with it.

%In Figure \ref{fig:DESphotoz} we plot the ratio of the density centred at the group redshift to the average of the density in the control apertures as a function of the group ID. The uncertainty on this quantity increases with redshift (here group IDs are sorted by redshift) as a result of the lower number of galaxies in the DES catalogue at higher redshift and of their more uncertain photo-z. 
Despite the limitations of this approach, we do not find a significant over-density of bright galaxies over a scale larger than the MUDF footprint for any of the groups studied. We therefore conclude that our inferred group properties are broadly representative of the underlying population.  Nevertheless, wider-area multi-object spectroscopic observations are required to investigate with better precision the large scale environment of the MUDF.

%\begin{figure}
%    \centering
%    \includegraphics[width=0.48\textwidth]{MUDF_DES_photoz.pdf}
%    \caption{Ratio of the galaxy density 
%    in an aperture of 1 Mpc radius and within $\Delta v = \pm 5000$\kms of the redshift of the groups to a control aperture of the same size and velocity depth but offset in redshift from the previous aperture and non overlapping with it. Densities are obtained as the sum of the fraction of the DES photo-z PDF within the reference aperture. A significant over-density of bright galaxies over a scale larger than the MUDF footprint is not found for any of the groups studied.}
%    \label{fig:DESphotoz}
%\end{figure}

\section{The CGM of group galaxies} \label{sec_mgiiabs}

Having completed a deep census of galaxy groups including members that are as faint as $m_{\rm MUSE} \approx 28.5$ mag (or stellar masses down to $\approx 10^{8}~\rm M_*$ at $z=1.5$), we now take advantage of the presence of the two bright quasars in the field to probe the gas content of galaxies within these structures in absorption. 
In particular, we focus on the \MgII\ absorbers detected in the quasar spectra at $z<1.5$. 

\subsection{\MgII\ fitting in the high-resolution quasar spectra}

To identify and fit the \MgII\ absorbers in the quasar spectra we employed a two step procedure. First, strong absorption lines were searched for by visually inspecting the spectra { across the entire wavelength range}, looking first for the \MgII\ doublets and then for other transitions which are commonly detected in quasar spectra, e.g. \MgI, \FeII,  \MnII, \CrII\ and \CaII. This procedure has been repeated by two authors independently (MFo, VDO) and for both quasars. We reliably identified five absorbers in the bright quasar at $z\approx 0.67,0.88,0.98,1.05,1.15$, while in the fainter quasar we only identify one absorber at $z\sim0.88$.

For each identified absorber we then fit the \MgII\ doublet profile with a novel method that uses Bayesian statistics to identify the minimum number of Voigt components required to model the data. Our method has two desirable properties: first, it requires minimal input from the user as no initial guesses are required; second, the final result is an optimal statistical description of the data, as the fit does not depend on any particular choice of initial guesses nor on a user-defined prior on the number of components.    
The details of the algorithm used will be presented in a forthcoming publication, and are only briefly summarised here. 

\begin{figure*}
    \centering
    \includegraphics[width=1.00\textwidth]{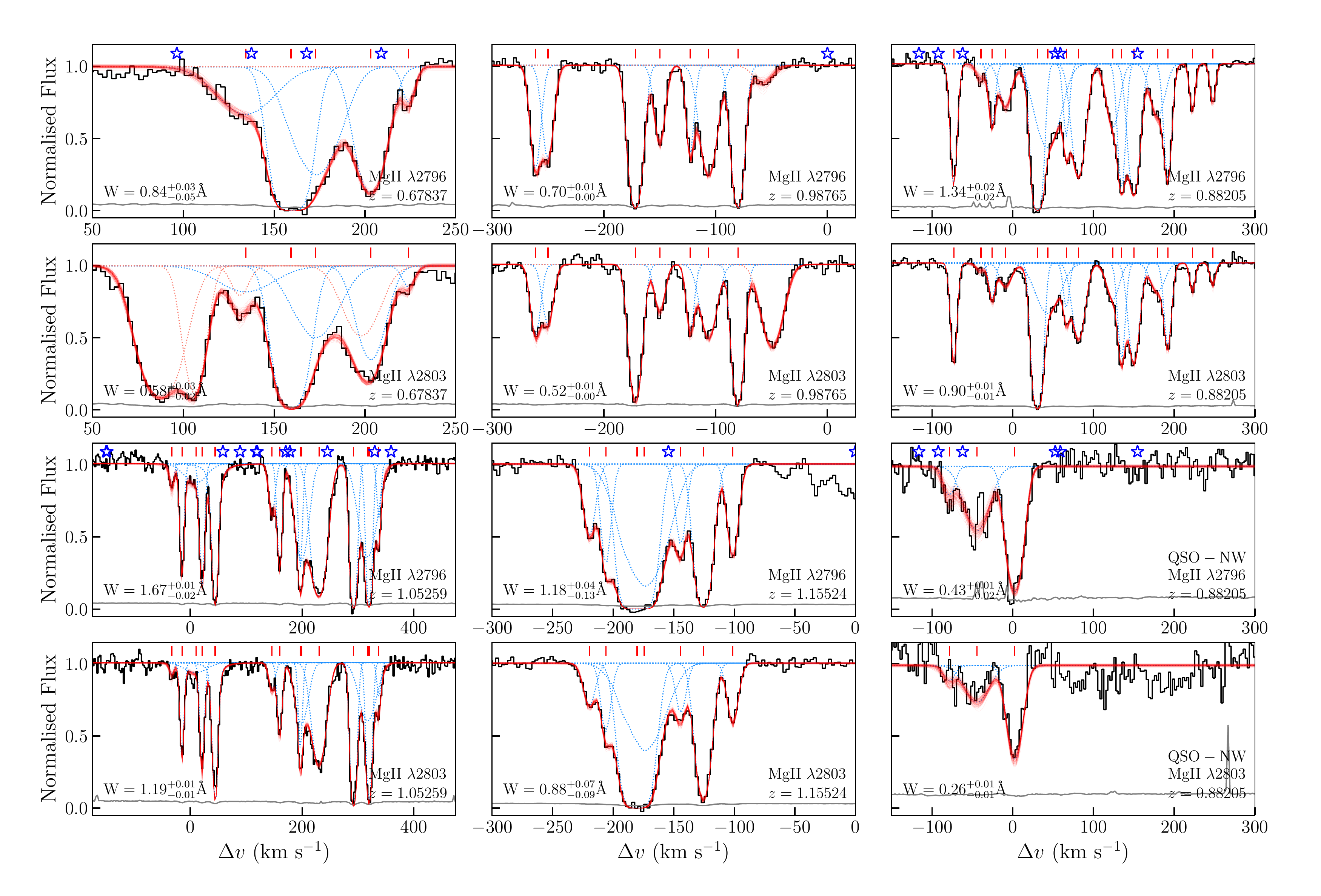}
    \caption{UVES spectra of \MgII\ absorbers identified in the brighter \qsoone\ (all panels except last two rows in the rightmost column), and in the fainter \qsotwo\ (last two rows in the rightmost column). The zero velocity corresponds to the average redshift of the galaxy group associated with the absorber, or to the redshift of the closest galaxy for the $z\approx 0.98$ absorber. The observed spectrum is shown as a black histogram with the $1\sigma$ error array in grey. A Voigt profile Bayesian fitting gives the red models, where individual lines are sampled from the posterior distribution. We include the rest-frame total equivalent width measurements in each panel. The red ticks above the continuum level show the position of individual \MgII\ absorption components, which are shown as cyan dotted lines. Pink dotted lines are for filler components arising from blends of different transitions. The blue stars show the spectroscopic redshift of MUDF galaxies.}
    \label{fig:fits_briqso}
\end{figure*}

Once the atomic transition (or transitions in case of multiplets to be fit jointly) has been identified, it is only required to specify the wavelength range (or disjoint ranges) to fit the normalized spectrum. The code then performs a fit of the data with progressively higher number of Voigt components (between a minimum and a maximum number), where each component is defined by a column density ($N$), { a Doppler parameter ($b$), and a redshift ($z$)}. For our fitting of \MgII\ absorbers we define uniform priors on the first two quantities in the range $11.5 < \log(N/{\rm cm^2})< 16.0$ and $1<b/{\rm km~s^{-1}}<30$, chosen to avoid modelling features within the noise or large-scale variations in the continuum level. The prior range on redshift is instead internally defined such that the absorption components can cover the selected wavelength range. A constant continuum level is also left as a free parameter in the fit to account for slight imperfections in the normalisation of the spectra. If required following a visual assessment of the doublet profiles, we allow for { a user defined number of} ``filler'' Voigt profiles designed to describe absorption lines arising from blends of different ions at different redshifts. One example is the rightmost line in the profile of the $z\approx0.98$ absorber in Figure \ref{fig:fits_briqso}. The model spectrum is then convolved to match the line spread function of the observed data using a Gaussian kernel. 

At each iteration, having defined a model with $n$ components, we sample the multidimensional likelihood space using {\sc PolyChord} \citep{Handley15}, a nested sampling algorithm that has better performance than {\sc MultiNest} for high-dimensional parameter spaces with multiple degeneracies between parameters. This is indeed our case, where complex models can reach a number of free parameters in excess of 50. After the sampling, we use the posterior samples to derive the Akaike Information Criterion \citep[AIC,][]{Akaike74} of each fit, defined as AIC $= 2\times n_{\rm par} - 2\times {\rm ln}(L)$, where $n_{\rm par}$ is the number of free parameters and $L$ is the likelihood of the model. We select as optimal fits those with a likelihood ratio within 1:150 compared to the best {(lowest)} AIC. If the selection includes only one model this is used as our best fit, otherwise we choose (from the selected ones) the model with the minimum number of components. 

Figure \ref{fig:fits_briqso} shows the fits of the \MgII\ absorption systems identified in the quasar spectra. While the fits are jointly obtained from the 2796~\AA\ and 2803~\AA\ transitions, we show only the stronger 2796~\AA\ transition in this figure. The single absorber found in the fainter quasar, which we discuss in more detail in Section \ref{sec_correlatedabs}, is shown in the Bottom Right panel, while in the other panels we show the absorbers found in the bright quasar sightline. In all panels we find complex kinematic profiles requiring from 3 to 15 Voigt components. Moreover we note that, with the exception of only one system ($z\sim0.98$), the absorbers are found at the redshift of a galaxy group identified in the previous section. Therefore, we set the zero of the velocity axis at the redshift of the galaxy group, except for one absorber where it is set to the redshift of the closest galaxy (ID 14) in velocity space. We stress that the position of the zero on the velocity axes is only made for reference and it does not affect our results. 

\subsection{Connecting the absorption features and the galaxy population} \label{sec_connectingprop}
The fact that \MgII\ absorbers are preferentially found at the redshift of a galaxy group ($5/6$ cases) suggests a direct connection between overdensities of galaxies and large cross sections of cold gas. Leveraging the deep spectroscopic survey in the MUDF that is complete to faint levels/low masses, we systematically explore this connection, in order to understand the physical origin of the gas probed in absorption. 

We show in Figure \ref{fig:fits_briqso} the spectroscopic redshift of the galaxies in the MUDF which are within the plotted range for each absorber (blue stars). We immediately note two features: first, each absorber (with the exception of the one at $z\approx0.98$) has one or more galaxies which are overlapping in velocity space with the absorption profile.
%, including several cases where at least one galaxy is aligned in velocity space with strong absorption components. 
Second, the absorbers with more complex kinematics ($z\sim0.88$, and $z\sim1.05$) have the largest number of galaxies at a small velocity offset. The case of the $z\sim0.98$ absorber, instead, stands out for having a complex profile, yet a single galaxy which is offset in velocity. However, given that the bright quasar is close to the South-East edge of the observed field, we cannot rule out the presence of other objects, just outside the MUDF footprint but at smaller velocity separation compared to the galaxy we found. Another issue is that the detection of continuum sources is more difficult at small projected separation from the quasars due to their brightness. We can detect sources as faint as $m_{\rm MUSE} \simeq 27~\rm mag$ at $\approx 25$ kpc and $\approx 18$ kpc in projection from \qsoone\ and \qsotwo, respectively. However, brighter or highly star forming galaxies would be detected even at smaller projected distances.

Thanks to the integral-field spectroscopic coverage of the MUDF, we next investigate trends in the spatial location of the galaxies which are found in the velocity window defined by the absorbers. Starting with the $z\sim0.67$ absorber as a first example, we note how the \MgII\ profile spans a velocity range $\Delta v \sim 100-200$ \kms\ with respect to the redshift of the group. We recall that the latter quantity is the average of the redshift of the group galaxies, and so we should expect a roughly equal number of them above and below the average value.  However, not all of them need to be distributed isotropically with respect to the quasars. 
In fact, by looking at Figure \ref{fig:groups_gallery}, we find that the four galaxies with velocities overlapping with the \MgII\ absorber are typically closer to the bright quasar than the remaining galaxies in the group. Indeed, their average projected distance from \qsoone\ is $\approx 130$~kpc, while for the other galaxies it is twice as large, at $\approx 250$~kpc.

Despite the large uncertainties driven by small number statistics, a similar trend emerges from the analysis of the other absorbers associated with the groups, as summarised in Table \ref{tab:dist_grp_gal}. This table reports the typical distance from the sightline \qsoone\ of galaxies that overlap in velocity with the absorption profile ($d_{\rm in,win}$), compared to the distance for the remaining galaxies in the group. In all cases, galaxies overlapping in velocity with the absorption profile lie at closer  projected separation (by a factor of $\approx 1.5-2$) than galaxies that are not overlapping with the \MgII\ absorbers. 

Considering instead the groups with no \MgII\ association (ID 2, 3, and 7), Figure~\ref{fig:groups_gallery} reveals a lack of galaxies in the immediate surroundings of the \qsoone\ line-of-sight, in line with the trend above. These results therefore suggest that the cold gas absorbers traced by \MgII\ arise preferentially from the CGM of one or multiple galaxies which are closer than $\approx 100$ kpc from the quasar line of sight. It should however be noted that the probability of group galaxies giving rise to \MgII\ is unlikely to be exclusively modulated by proximity to the line of sight, with covering factors playing a role. Indeed, Figure~\ref{fig:groups_gallery} shows also the presence of galaxies in close proximity to the \qsotwo\ line of sight (mostly from groups 3, 4, and 5) that do not necessarily give rise to absorption (see below).

\begin{table}
    \centering
    \begin{tabular}{cccc}
        \hline
        $z_{\rm abs}$ & Group ID & $d_{\rm in,win}$ $\rm{(kpc)}$ & $d_{\rm out,win}$ $\rm{(kpc)}$  \\
        \hline
        0.67 & 1 & $130\pm56$ & $251 \pm 25$ \\
        0.68 & 2 & - & $257 \pm 94$ \\
        0.78 & 3 & - & $363 \pm 92$ \\
        0.88 & 4 & $274\pm87$ & $522$ \\
        1.05 & 5 & $162\pm41$ & $467\pm29$\\
        1.15 & 6 & $195$& $321\pm101$\\
        1.22 & 7 & - & $322 \pm 83$ \\
    \end{tabular}
    \caption{Average projected distance of the galaxies which reside in groups at the same redshift of a \MgII\ absorber, for galaxies within the velocity window defined by the absorption profile, and outside it. { For groups without an \MgII\ detection the average projected distance of all galaxies is given in the last column.} Values without errors have only one object that satisfies the selection threshold. }
    \label{tab:dist_grp_gal}
\end{table}

The relation between the strength of \MgII\ absorbers and the galaxy distance has been studied extensively in the literature. For instance, \citet{Chen10} report a large statistical sample of galaxy-quasar pairs, finding a strong anti-correlation between the equivalent width of the absorption profile and the distance of the closest galaxy. 
We can therefore compare directly in this parameter space the properties of MUDF galaxies, both within groups and in isolation, with the results present in the literature.

\begin{figure*}
    \centering
    \includegraphics[width=0.98\textwidth]{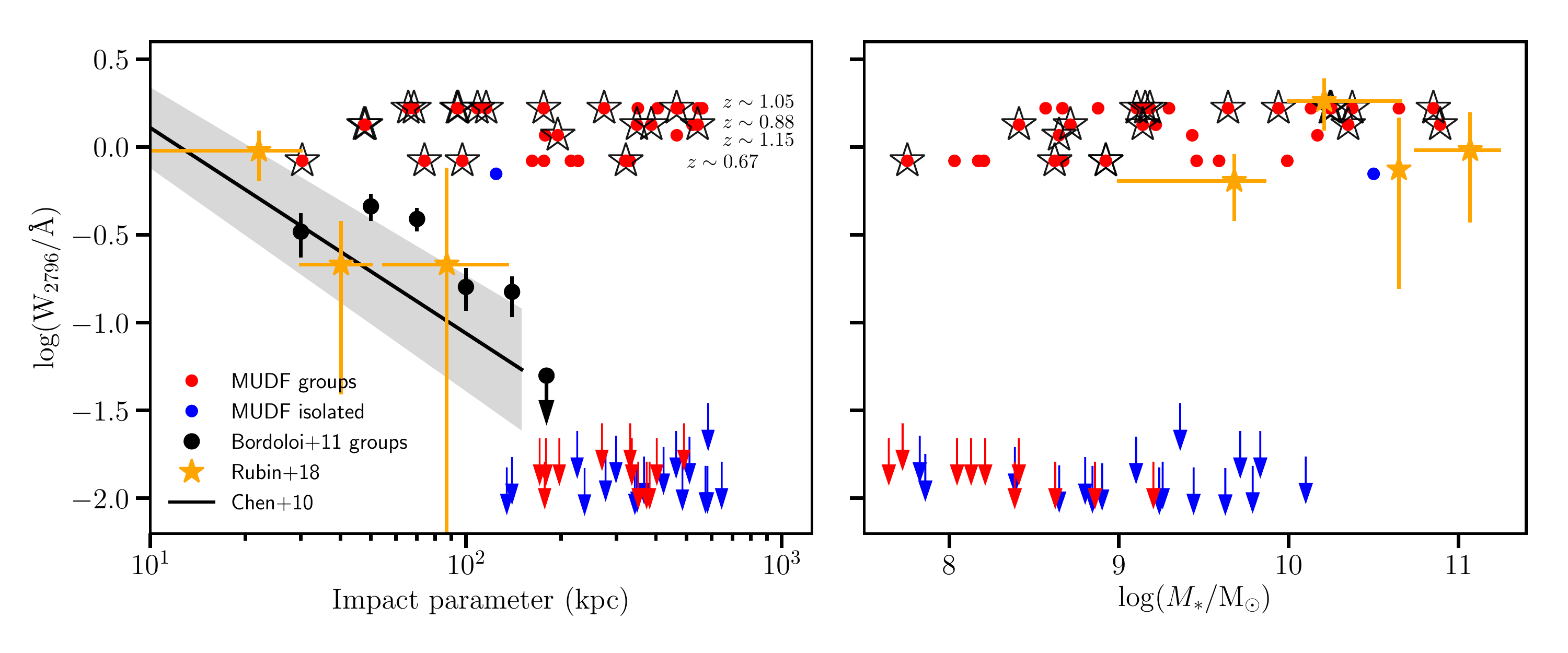}
    \caption{Rest-frame equivalent width ($W$) for the \MgII\ $\lambda2796\AA$ absorption line as a function of impact parameter { (projected distance of each galaxy from the \qsoone\ sightline for MUDF galaxies, left panel)} and of the galaxy stellar mass (right panel). Galaxies in the MUDF footprint are shown as red circles if they reside in a group and as blue circles if they are isolated. The arrows (red for galaxies in groups and blue for isolated galaxies) show $2\sigma$ upper limits measured on the UVES spectrum in a 10 \kms\ velocity window centred at the redshift of each galaxy. The black empty stars show galaxies which have a redshift within the \MgII\ absorption profile for the strong absorbers shown in Figure \ref{fig:fits_briqso}. The redshift of the groups are shown in the left panel at the y-axis levels of the corresponding values of $W$.  The black line is the best fit relation (with the 1-$\sigma$ confidence area shaded in grey) between $W$ and impact parameter from \citet{Chen10}. The orange stars { and black circles are obtained from stacking galaxy-galaxy pairs from the sample of isolated galaxies of \citet{Rubin18} and the sample of group galaxies of \citet{Bordoloi11}.}}
    \label{fig:mgii_w_dist_mass}
\end{figure*}

In Figure \ref{fig:mgii_w_dist_mass}, we show the rest-frame equivalent width ($W$) for the \MgII\ $\lambda2796\AA$ absorption line, obtained integrating the { best-fit} models derived above, as a function of the projected distance of each galaxy from the \qsoone\ sightline (left panel) and of its stellar mass (right panel). Galaxies belonging to a group that is associated with an \MgII\ absorber are shown as red circles and they have been assigned the total $W$ of the absorption profile, while the galaxy which we associate with the $z\sim0.98$ absorption system is shown as a blue circle. For the other sources, we plot $2\sigma$ upper limits measured on the UVES spectrum in a 10 \kms\ velocity window centred at the redshift of the galaxy. 

From this analysis, it appears that the 4 out of 7 groups that are associated with an absorber have high equivalent width and, as noted above, have at least one galaxy which is within 150~kpc of the quasar line of sight.  Moreover, these groups tend to host galaxies that are relatively massive ($\log(M_*/\rm{M_\odot})>10$) and therefore, given their CGM is likely to scale with their size \citep[e.g.][]{Chen10}, they are more likely to have a larger cross section of \MgII\ that can give rise to the absorption. 

An apparent exception is the $z\approx0.67$ group, which seems to lack particularly massive galaxies. However, we argue that in this case the \MgII\ absorption is mostly driven by the fact that one galaxy, despite being of a lower stellar mass ($\log(M*_/\rm{M_\odot}) = 8.92$), is very close in projection (29~kpc) and its redshift places it at the centre of the absorption profile.  On the other hand, for the three groups without a detection, their galaxies are both further away from the absorber and their stellar masses are low $\log(M_*/M_\odot)\sim 9$, overall reducing the chances that the CGM of group members intercepts the line of sight. 

This picture is further reinforced when we make the same analysis for the \qsotwo. We find that the galaxies in groups are on average further away from this line of sight compared to that of the brighter quasar. Only three galaxies are within 100 kpc, two of which are in the $z\approx0.88$ group which we associate with the only \MgII\ detection in this sightline (see Section \ref{sec_correlatedabs}).
Moreover, a third lower-redshift quasar ($z\approx 1.285$; cyan cross in Fig. \ref{fig:groups_gallery}), lying close to the edge of the FoV (where no sources have been extracted) is in close spatial proximity to the $z\approx$0.67 group. It does in fact exhibit strong \MgII\ absorption at this redshift as seen from the MUSE spectrum, but { we did not find} other absorbers associated with groups or individual galaxies at larger separations from this third sightline. 
We therefore conclude that a combination of proximity to the quasar line of sight and presence of massive galaxies hosting large cross sections of cool gas are the main factors that control the high incidence of \MgII\ absorbers in these groups.

Having established what drives the incidence of \MgII\ in these groups, we now compare the absorption properties to samples of galaxies from the literature. To this end, in Figure \ref{fig:mgii_w_dist_mass}, we show the results of the analysis by \citet{Chen10} (black solid line) and the average properties of the CGM of galaxies at $0.35 < z < 0.8$ from foreground and background galaxy pairs in the PRIMUS survey \citep{Rubin18}. These galaxies are not explicitly selected to be within groups. These authors find that the strength of \MgII\ absorption declines as a function of impact parameter ($d$, i.e. the projected separation of a galaxy from the probing sightline), with the average gas content of star forming galaxies versus $d$ in the PRIMUS survey being consistent with the fitting function of \citet{Chen10}, albeit with a large intrinsic scatter for individual galaxies. 

However, the total equivalent width of \MgII\ in galaxies within the MUDF groups is higher than the $1\sigma$ scatter of the data from \citet{Rubin18} when shown as a function of $d$, but are consistent with these data as a function of stellar mass. Likewise, compared to the scaling relation of \citet{Chen10}, the MUDF group galaxies lie consistently above the relation at fixed impact parameter. 
These results imply that the group environment has an effect on the cross section of \MgII, leading to enhanced equivalent widths at larger distances compared to galaxies not explicitly selected within groups. { \citet{Bordoloi11} presented the radial \MgII\ absorption profile from a sample of group galaxies from the zCOSMOS survey finding that groups have more extended absorption profiles compared to a sample of isolated galaxies from the same survey. However, the radial profile for group galaxies is largely consistent with the one from the \citet{Chen10} and \citet{Rubin18} samples. This could be an intrinsic feature of the sample or an effect of contamination from more isolated galaxies in the group sample. }\citet{Nielsen18} studied the \MgII\ absorption in a sample of 29 group-like environments (defined to have at least two galaxies within 200 kpc of a background quasar and with $\Delta v < 500$ \kms). These authors found that the \MgII\ median equivalent width in this sample is $W=0.65\AA$, and it is indeed enhanced by a factor 2.2 compared to an isolated galaxy sample. This is also in line with the result of \citet{Gauthier13}, who reported on the \MgII\ equivalent width for three galaxy groups associated with ultra-strong \MgII\ absorbers ($W>3.5\AA$), which are atypical of the general galaxy population. 
%None of our groups, however, are associated with such a strong absorption system, suggesting that these extreme equivalent widths are a rare occurrence. 

Finally, we comment on the properties of the only absorber associated with an isolated galaxy.  The complete isolation of this source (ID 14) is unlikely, given its stellar mass ($\log(M_*/M_\odot) = 10.51$). It is possibly living in a massive halo which in turn is expected to host satellite galaxies (see below). And while this source is the closest to the quasar line of sight among individual galaxies in the MUDF, its equivalent width is still elevated for its impact parameter. Moreover, the several narrow kinematic components found in the absorption spectrum seem to suggest, in analogy with the other \MgII\ arising from groups, that this absorber could be associated with a group that we do not fully cover in the MUSE footprint. 

\subsection{Detection of cold gas absorption in the quasar pair} \label{sec_correlatedabs}

Up to this point, our analysis reveals that groups hosting \MgII\ absorbers show galaxies in closer proximity to the line of sight and host more massive galaxies, two factors that are likely to boost the incidence of \MgII\ absorption. Moreover, the equivalent width of \MgII\ appears to be elevated compared to samples not explicitly selected in groups. We now fully exploit the ability to conduct tomography in the MUDF to assess whether the observed absorbers can be attributed to a widespread intra-group medium, or whether they are more likely to be associated with individual galaxies. 

The groups at $z\approx 0.88$ and $z\approx 1.05$ appear to be most suited for this analysis as they both contain a significant number of members that appear to be distributed across the two quasar sightlines. 
For the $z\approx 1.05$ absorber, despite a strong ($W=1.67 \AA$) and complex kinematic profile seen against the \qsoone, we derive only a 2$\sigma$ upper limit of 0.06\AA\ against the \qsotwo. For the group at $z\approx 0.88$, instead, we detect also  \MgII\ absorption in the UVES spectrum of the fainter \qsotwo. At this redshift, we find a \MgII\ absorber also in the spectrum of the bright \qsoone, which enables us to study whether the kinematics of the absorption is correlated over a scale of 500 kpc. The right panels in Figure \ref{fig:fits_briqso} show the data and Voigt profile models of the absorption systems found at $z\approx 0.88$ in both sightlines (top panel for the bright quasar and bottom panel for the fainter one). Along the line of sight of the faint quasar we find only three Voigt components, as opposed to 15 in the other sightline. These components span a much smaller velocity range ($\sim 100$ \kms) compared to the absorption system found in the brighter quasar ($\sim 350$ \kms). Furthermore, we { do not find a significant correlation between the kinematics of the components in the two sightlines. A marginal amount of correlation could be present in the first and last (in velocity space) components seen in the \qsotwo\ sightline, where a component exists at a similar velocity also in the other sightline, however with a significantly different strength.} This result suggests that the cold absorbing gas is not necessarily related to coherent structures in the group as a whole, which is also corroborated by the non-detection of correlated \MgII\ in the $z\approx 1.05$ group.

The lack of evidence in support of a dense and homogeneous intra-group medium leaves open the hypothesis that the enhanced absorption in the groups arises from the CGM of individual galaxies, 
which has been processed by the environment. Indeed, considering again the 
$z\approx 0.88$ group, we find near the \qsotwo\ line of sight  a clustering of galaxies with negative velocities compared to the group average redshift, which broadly corresponds to the velocities of the absorption components in this sightline. More specifically, from Figure \ref{fig:groups_gallery}, it appears that the galaxies lie in two sub-groups that are mostly aligned along the SE-NW direction. The fact that both quasar sightlines go through the direction of this alignment is a possible explanation for why this group is the only one detected in both sightlines. 

%\begin{figure}
%    \centering
%    \includegraphics[width=0.45\textwidth]{MUDFpc_correlated_MgII.pdf}
%    \caption{UVES spectra of the \MgII $\lambda$2796 absorber at $z\approx0.88$ from the sightline of the bright quasar (\qsoone, top panel) and from the sightline of the faint quasar (\qsotwo, bottom panel). The zero of the velocity scale corresponds to the average redshift of the galaxy group associated with the absorber. The lines and symbols are coded as described in Figure \ref{fig:fits_briqso}.}
%    \label{fig:fits_correlated}
%\end{figure}

\begin{figure}
    \centering
    \includegraphics[width=0.48\textwidth]{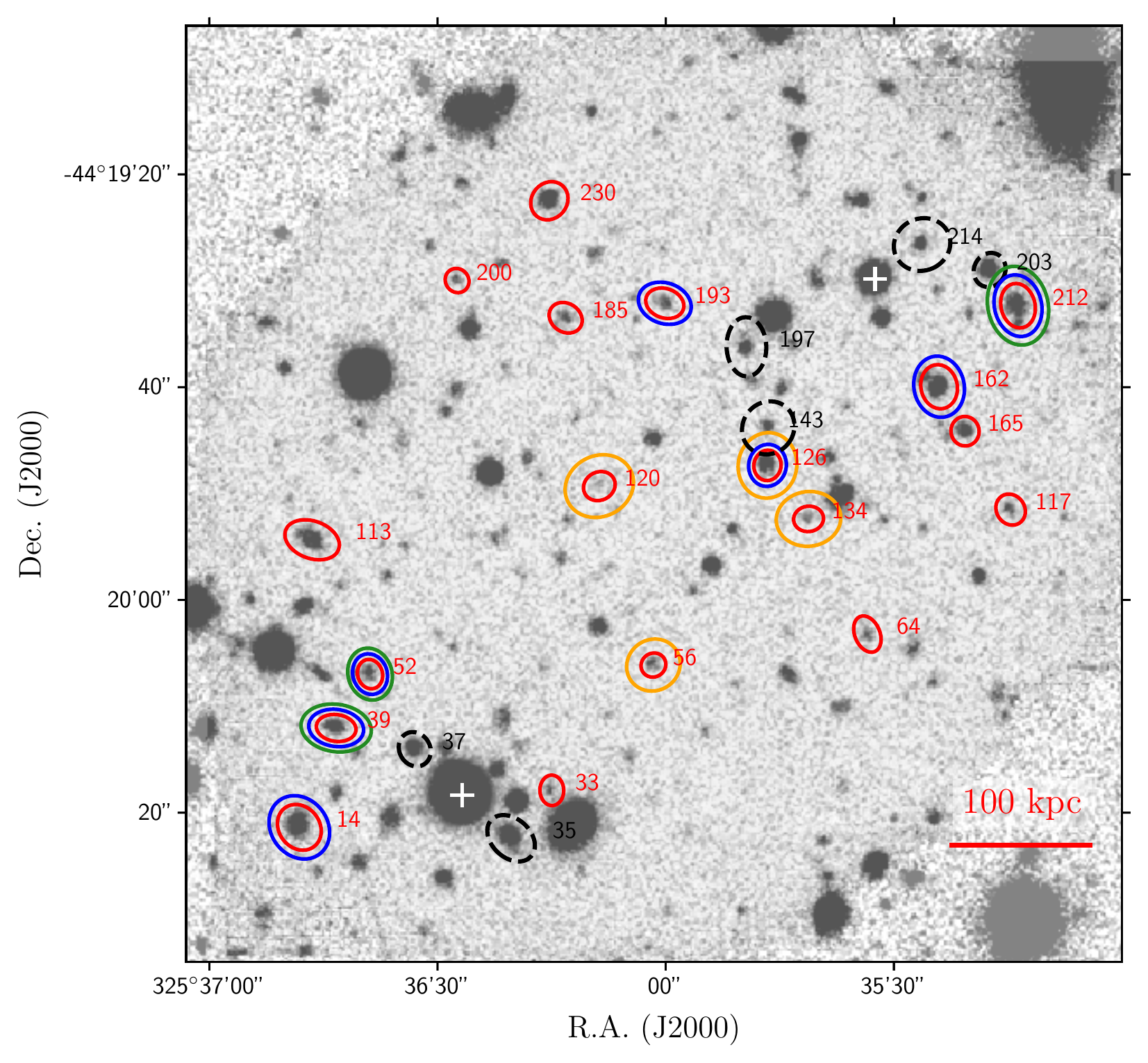}
    \caption{ Location within the MUDF of the background galaxies (red solid ellipses) that we include in the spectral stack A used to search for extended \MgII\ absorption around the galaxies (black dashed ellipses) in the  $z\sim0.88$ group. Three more stacks (B, C, and D) include the galaxies marked with blue, green, and orange ellipses respectively.}
    \label{fig:musefovstack}
\end{figure}

\begin{figure}
    \centering
    \includegraphics[width=0.48\textwidth]{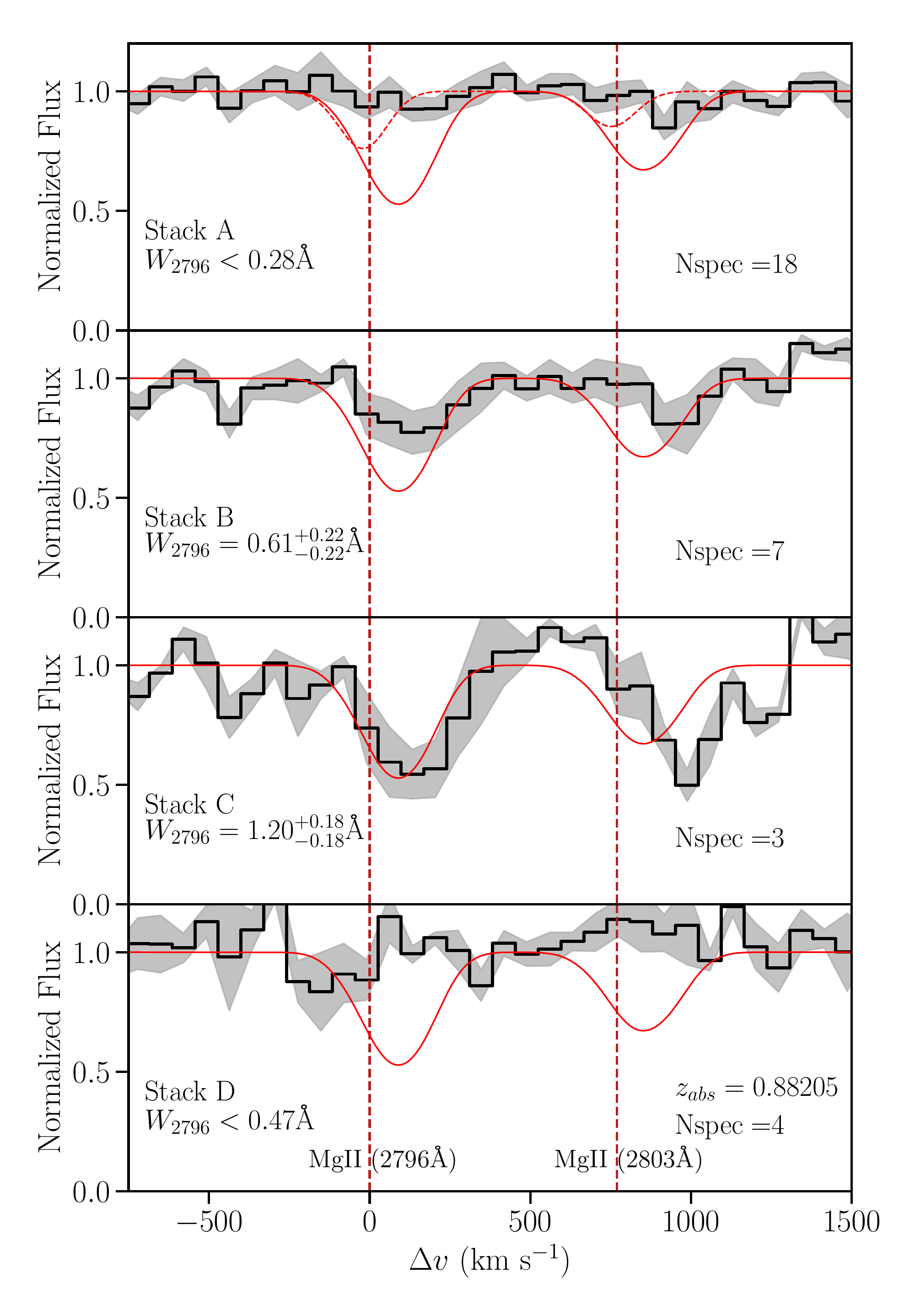}
    \caption{Stacked spectra (black) of MUDF background sightlines probing galaxies of the $z\sim0.88$ group around the \MgII\ doublet. { The panels from top to bottom refer to stacks A to D as described in the main text.} The gray filled region shows the 1$\sigma$ uncertainty from bootstrap resampling. The zero velocity is defined by the average redshift of the group ($z=0.88205$), and the rest velocity of each transition in the \MgII\ doublet is shown by red vertical dashed lines. The red solid line shows the best fit model of the UVES spectrum of the QSO-SE convolved to match the MUSE spectral resolution, { while the red dashed line in the top panel shows the best fit model for the QSO-NW spectrum.}}
    \label{fig:stack7}
\end{figure}

Thanks to the very deep spectroscopy in the MUDF we can further test for the presence of a diffuse cool intra-group medium by stacking the spectra of background galaxies that lie behind the $z\approx 0.88$ group. { We analyse four stacks: A) 18 galaxies selected to have a $S/N>3$ per wavelength channel in two spectral windows bracketing the wavelength of the $z\approx 0.88$ absorption (red ellipses in Figure \ref{fig:musefovstack}); B) seven galaxies that are roughly co-spatial with the distribution of the group galaxies (blue ellipses); C) three galaxies which show \MgII\ absorption at $z\approx 0.88$ in their individual spectra (green ellipses); and D) four galaxies lying between the two QSO sightlines and roughly aligned with the geometrical center of the group (orange ellipses).}
For comparison, the position of the group galaxies is shown by black dashed ellipses. 
The stacked spectra are shown in Figure \ref{fig:stack7} as a black lines, with 1$\sigma$ uncertainty from bootstrap resampling shown in grey. 

{ In stacks A and D we find non detections with 2$\sigma$ upper limits of $W_{2796}<0.28\AA$ and $W_{2796}<0.47\AA$ respectively. In stack B, which is restricted to the galaxies aligned with the two foreground sub-groups, we find a $\sim3.5\sigma$ combined detection of the \MgII\ doublet with $W_{2796}=0.61^{+0.22}_{-0.23}\AA$, although with a $\approx 50$ \kms\ offset with respect to the absorber seen in the bright quasar sightline (red solid line). We checked by comparing the MUSE and UVES spectra of the quasars that this offset is real, and does not arise from mismatches in the wavelength calibration. Lastly, in stack C, which is composed of galaxies in which the absorption signal can be detected in individual spectra, this detection becomes even stronger with $W_{2796}=1.20^{+0.24}_{-0.21}\AA$.
Again, this result reinforces the idea that there exists a large cross section of cool gas in close proximity to the group galaxies (stacks B and C), but that there is no dense and widespread cool gas that gives rise to strong absorption signal filling larger scales beyond the regions traced by galaxies themselves (stack A), or the region between the two sub-groups (stack D).} 

\section{Discussion} \label{sec_discussion}

\subsection{A high fraction of strong \MgII\ absorbers in galaxy groups}
We performed a search for galaxy groups at $0.5<z<1.5$ in the MUDF, 
finding seven groups of at least three galaxies. From abundance matching we obtain a virial mass for these groups in the range $\log(M_h/\rm{M_\odot}) \approx 11 - 13.5$, with virial radii in the range of $\approx 100-200~\rm kpc$. Our most massive group includes 15 galaxies and could evolve into a Virgo-like cluster of galaxies by $z=0$ \citep{Chiang13}. 

An independent search for cool gas absorbers, as traced by the \MgII\ doublet in the same redshift range, revealed five absorption systems in the high resolution spectrum of the brighter \qsoone, and one system in the spectrum of the fainter \qsotwo. In total, of these 6 absorbers, 5 are at the same redshift as a galaxy group, a fraction of 83$\%$. For the only absorber not associated with a group, we found a galaxy within 300 \kms\ in velocity space. This galaxy is massive ($\log(M_*/M_\odot) = 10.51$), which correponds to a dark matter halo mass of $\log(M_h/M_\odot) \approx 12.25$, following \citet{Moster10}. At this mass, it is very likely that the halo hosts several satellite galaxies which could be outside our survey footprint, so with spectroscopic data covering a larger area we might find this galaxy to be part of a group as well. Indeed, by looking at the absorption profiles in Figure \ref{fig:fits_briqso}, this absorber looks kinematically complex, with several strong and narrow absorption components as found for the absorbers associated with galaxy groups. 

Despite the alignment in redshift space between the group galaxies and the absorption profiles, it is difficult to unambiguously conclude whether we are observing gas in the intra-group medium, or whether 
absorption rises ultimately from the CGM of individual group members.
Throughout our analysis, however, we have uncovered pieces of evidence that more closely support the latter scenario.

We have found that all the \MgII\ detections arise in groups with at least one galaxy closer than 150 kpc from the bright quasar line of sight, and the same is true for the single absorber in the faint quasar.
Moreover, the group galaxies which are closer in redshift space to the absorbers are also closer in projection, 
and groups with a \MgII\ detection appear to host relatively massive galaxies that are expected to have intrinsically larger cross sections of cool gas in their CGM. Similarly, all the non-detections are associated with galaxies at larger distances from the sightline, or that are at lower mass (and hence likely have smaller cross sections) compared to the members of groups with a \MgII\ detection. 

We have not found compelling evidence for the presence of a widespread intra-group medium { producing strong absorption} for two reasons. First, out of four \MgII\ systems associated with groups, only one system (at $z\approx 0.88$) is detected in the two sightlines, and the profiles are not very well correlated in velocity space. Second, when stacking galaxies in the background of this $z\approx 0.88$ group, a detection of \MgII\ in absorption emerges only when considering background galaxies in at small projected separations to the foreground ones.  Altogether, these pieces of evidence suggest that the { strong absorbers} are associated mostly with the CGM of group galaxies. { However, we do not rule out the presence of a homogeneous low-density intra-group medium which can give rise to weak absorption components.}

This conclusion, however, does not rule out environmental effects from shaping the properties of the CGM of group members. Indeed, when considered by their equivalent width, group galaxies appear to host stronger \MgII\ absorbers compared to the general galaxy population.  Several studies that have focused on samples of (relatively massive) galaxies probed by quasar sightlines, not selected according to their environment, have uncovered a clear  anticorrelation between $W$ and impact parameter $d$ \citep{Steidel94, Chen10}. A similar trend seems to hold also in samples where background galaxies are used to probe foreground ones \citep{Rubin18}.
In these samples, strong absorbers with equivalent width $W>1\AA$ are typically found at $d<30$ kpc. Conversely, in our groups, strong systems are found at distances $\approx 50-100~\rm kpc$, despite the fact that we do not necessarily probe more massive galaxies than these previous studies (see Figure \ref{fig:mgii_w_dist_mass}). 

While the most massive galaxies ($M_* \approx 10^{11}~\rm M_\odot$) in samples from the literature are themselves likely members of groups, generally these samples are thought to be more representative of the field population. Indeed, using a light cone from the \citet{Henriques15} semi-analytic model of galaxy formation, we find that for galaxies with a stellar mass in the range $10^{9.75}-10^{10.25}{\rm M_\odot}$ and redshift $0.4<z<0.6$, a fraction of only 42\% live in groups as defined by our method, with this fraction reducing even further at lower stellar mass. Thus, the samples of \citet{Chen10} and \citet{Rubin18}, which in some cases extend down to $10^9~\rm M_\odot$, are representative of a more isolated population of galaxies compared to the sample of group galaxies studied in this work. 
Therefore, the difference in absorption strength at fixed impact parameter, combined with the complex kinematical profiles, imply that the group environment plays an active role in boosting the cross section of cool gas present in the CGM of member galaxies.

This result, that emerges from a complete and systematic search of groups in the MUDF, is in line with previous findings from serendipitous discoveries of groups associated with \MgII\ absorbers.  
For instance, \citet{Whiting06} found at least five galaxies at the redshift of a $W=2.5\AA$ \MgII\ absorber. Similarly, \citet{Kacprzak10} found five low-mass galaxies associated with a $W=1.8\AA$ absorber. \citet{Nestor11} found ultra-strong \MgII\ absorbers with $W=3.6, 5.6 \AA$ in two systems associated with massive galaxy pairs, and lastly \citet{Gauthier13} found another very strong absorber with $W=4.2 \AA$ in a group of three galaxies. These works found that the closest galaxy is at $d=30-70$ kpc. In the work by \citet{Kacprzak10} a galaxy is found a smaller impact parameter (17.4 kpc).  However, from an analysis of the metallicity of the absorber and of the group galaxies, the authors concluded that the closest galaxy is not hosting the absorber, which is more likely to arise from tidal debris in the group environment. { \citet{Bordoloi11} found marginally more extended \MgII\ absorption profiles compared to a sample of more isolated galaxies, concluding that the absorbing gas is more likely to be associated to the individual galaxies rather than the intragroup medium}. \citet{Nielsen18} similarly found an enhanced equivalent width of \MgII\ absorbers in a sample of 29 galaxy groups with respect to an isolated galaxy sample. These authors have complemented these results with a kinematical analysis of the absorption profiles and concluded that the absorbers originate from an intragroup medium rather than from individual galaxies. However, it remains unclear if these sparse groups (having on average 2-3 members) are part of a virialized halo that is able to host a diffuse intragroup medium. { Lastly, \citet{Bielby17a} studied the \MgII\ absorption profile associated to a low-mass group at $z\approx0.28$ finding that the absorbing gas is likely to arise from multiple gas clouds orbiting in the group halo and giving rise to the intragroup medium.}

%Lastly, \citet{Klitsch18}, combined ALMA and MUSE data to study the properties of an \HI\ absorber in a group at $z\approx0.63$, concluding that the absorber is associated with gas in the intragroup medium. }

\subsection{The origin of enhanced cold gas in galaxy groups}

We have uncovered evidence supporting a scenario in which, at fixed stellar mass, group galaxies have a larger cross section of cool gas that gives rise to frequent and strong absorption systems. A question arises about what mechanisms may be responsible for this enhancement.

Several physical mechanisms might contribute to a larger cross section of \MgII\ in group galaxies, including higher gas fractions for satellite galaxies in dense environments \citep{Noble17} or stronger outflows from stellar winds due to enhanced star-formation \citep{Mcgee14}.  There seems however to be no clear consensus on the relevance of these mechanisms in groups at $z\approx1$ \citep{Wetzel12, Rudnick17, Fossati17}. Moreover we find it difficult to assess the contribution of these mechanisms with our own data. As we argue below, however, a prominent mechanism at play is likely to be gravitational interaction among group members.

It is well known that galaxy groups are the ideal environment to trigger gravitational interactions among members. Indeed, in more massive haloes (i.e. galaxy clusters), the higher velocity dispersion leads to shorter interaction times, with a reduced effect of tidal interactions on the gaseous and stellar structure of galaxies \citep[see][]{Boselli06}.
Massive groups are therefore a sweet-spot in terms of richness and velocity dispersion for gravitational encounters to occur. 
For this reason, we argue that the evidence described above suggests that absorbers arise from gas once bound to the group galaxies (or their CGM), which has been stripped by tidal forces. This material, displaced from its original site, naturally boosts the cross section of cool gas, leading to enhanced absorption at larger impact parameters compared to the field galaxies \citep{Morris94}. 
These tails, gaseous bridges, and plumes have indeed been imaged in atomic and ionized gas within groups in the local Universe \citep{Mihos12, Rasmussen12, Taylor14, Fossati19}, where the denser components can stretch up to scales of hundreds of kpc. In this picture, the several kinematic components seen in the absorbers could be related to the chaotic orbits of the group galaxies during the stripping process, as seen for instance in the complex kinematic maps of galaxy encounters traced in emission with MUSE \citep{Fossati19}.

This process would explain why a correlated detection of cool gas in both the MUDF quasars is a rare event. Given the distance of $\sim 500$ kpc between the two sightlines, a double detection requires either a very massive group that spans this distance with enough galaxies, a number of which is subject to some degree of gravitational perturbation. 
Alternatively, a special alignement of galaxies is required, as in the $z\sim 0.88$ group, which is composed of two sub-groups that fortuitously align with the orientation of the two quasars. 

The connection between enhanced \MgII\ absorption and a tidal stripping scenario is further reinforced by other works. For instance, \citet{Kacprzak10}, using the high-resolution of {\it HST} imaging data, found perturbed morphologies for the three brightest group galaxies, with tidal tails extending up to $\sim25$ kpc. They conclude that these morphological features are suggestive of merger events or tidal stripping and that dense and cool stripped gas can host the observed \MgII\ absorber. 
More recently, \citet{Chen19a} studied the same group with deep MUSE observations. These data corroborated the results of \citet{Kacprzak10} revealing a giant nebula of ionized gas contributing to the total mass and metal content of the intra-group medium. An accurate kinematical analysis showed that the \MgII\ absorber is indeed located in the stripped gas passing in front of the background quasar. These results point towards the presence of a multi-phase medium { \citep[see also][]{Bielby17a}}, and it is possible that the cool and ionized gas is gradually heated to the group virial temperature, contributing to the warm-hot intra-group medium.

\section{Summary and Conclusions} \label{sec_conclusions}
In this work, we have presented the design, observations, and data reduction methodology for the MUSE Ultra Deep Field (MUDF) survey, together with results from the cold gas content of galaxy groups at $0.5<z<1.5$. The MUDF survey is a 150-hour (on source) large programme on the MUSE instrument at the VLT that is observing a $1.5\times1.2$ arcmin$^2$ region of the sky characterised by the presence of two quasars at $z\approx 3.22$ separated by $\approx$ 60 arcsec. The MUSE data are also complemented by {\it HST} imaging programmes in the near-UV and in the near-IR, and by the deepest {\it HST} near-IR spectroscopic campaign in a single field. Deep high-resolution spectroscopy of the quasars is also being collected with the UVES instrument at the VLT. These rich datasets will enable us to reach several goals, including: an investigation into the connection between gas and galaxies reaching the low-mass regime at $z\sim 2-3$; a search for gas filaments of the cosmic web in emission; and a study of the build-up of the Hubble sequence with cosmic time with accurate information on the morphology, kinematics, gas budget, and star-formation history of galaxies.  

In this paper, we have discussed in detail the survey design, the data reduction procedure, the extraction and validation of source catalogues, and the derivation of the physical properties of galaxies from the partial dataset observed to date. As a first application, we investigated the galaxy environments in the MUDF, finding seven groups three or more members at $0.5<z<1.5$ covering a large range in inferred dark matter halo mass ($\log(M_h/\rm{M_\odot}) \approx 11 - 13.5$).
We explored the correlation between galaxies and galaxy groups with cold gas detected via \MgII\ absorption in the quasar sightlines. We found five absorption systems in the spectrum of the bright quasar and only one in the faint quasar spectrum. All but one of these systems are at the redshift of a galaxy group, while for the last one we find a massive galaxy at a small velocity separation, which we speculate could itself be a member of a group falling outside the MUDF footprint. 

The absorbers have a complex velocity profile which we decompose into several Voigt components using a novel Bayesian technique. We find that, within a given group of galaxies, members that are close in velocity space to the absorber are also closer to the sightline in projection compared to the other group galaxies. Furthermore, through the analysis of correlated absorption in both quasar sightlines and via a tomographic map of one of the groups using background galaxies,  we find no significant evidence of a widespread and homogeneous intra-group medium giving rise to { a strong} absorption signal.   
Altogether, these results suggest that the absorbers reside in (or are stripped from) the CGM of one or more galaxies within $\approx 100$ kpc from the quasar sightline. 

The strength of the absorption seen in groups is higher than what is typically found for more isolated galaxies at comparable impact parameters. This evidence, combined with previous examples in the literature of strong \MgII\ absorbers in group-like environments, suggests that the absorbers reside in gas once bound to individual galaxies (or their CGM) that has been stripped at larger radii, boosting the cross section of \MgII. Gravitational interactions and tidal forces, in analogy with what is seen in nearby groups, are indeed  effective in galaxy groups and obvious mechanisms to strip gas and stars from the galaxy disks, leading to the observed phenomenology. 

So far, our analysis relied on deep MUSE data in a double quasar field. These data will soon be complemented by deep spectroscopic and imaging data from {\it HST}, extending the wavelength of our observations from the near-UV to the near-IR. These data will provide us with accurate redshifts for galaxies in the redshift desert ($1.5 < z < 3.0$), as well as strong constraints on the stellar mass and recent star-formation of galaxies down to low stellar mass. The approach of combining multiple quasar sightlines with complete galaxy surveys and an accurate reconstruction of the galaxy environment is critical for understanding the role of strong absorbers in the framework of how galaxies are fed with fresh gas. 
In forthcoming papers, we will study the impact of the group environment on the star-formation activity of galaxies, while extending this work to higher redshift.

\section*{Acknowledgements}
We thank J.T. Mendel for the developement of the MC-SPF code used in this work and the anonymous referee for their comments which improved the quality of the manuscript.
M.Fumagalli acknowledges support by the Science and Technology Facilities Council [grant number  ST/P000541/1]. This project has received funding from the European Research Council (ERC) under the European Union's Horizon 2020 research and innovation programme (grant agreement No 757535). SC acknowledges support from Swiss National Science Foundation grant PP00P2\_163824. RC was supported by a Royal Society University Research Fellowship. CP thanks the Alexander von Humboldt Foundation for the granting of a Bessel Research Award held at MPA. MR acknowledges support by HST Program GO-15637 provided by NASA through grants from the Space Telescope Science Institute, which is operated by the Association of Universities for Research in Astronomy, Inc., under NASA contract NAS5-26555.
This work is based on observations collected at the European Organisation for Astronomical Research in the Southern Hemisphere under ESO programme IDs 65.O-0299(A), 68.A-0216(A), 69.A-0204(A), 1100.A-0528(A), 1100.A-0528(B), 1100.A-0528(C), 0102.A-0194(A), 0102.A-0194(B).
This work used the DiRAC Data Centric system at Durham University, operated by the Institute for Computational Cosmology on behalf of the STFC DiRAC HPC Facility (www.dirac.ac.uk). This equipment was funded by BIS National E-infrastructure capital grant ST/K00042X/1, STFC capital grants ST/H008519/1 and ST/K00087X/1, STFC DiRAC Operations grant ST/K003267/1 and Durham University. DiRAC is part of the National E-Infrastructure. This research made use of Astropy \citep{Astropy-Collaboration13}. For codes and data products used in this work, please contact the authors or visit \url{http://www.michelefumagalli.com/codes.html}. Raw data are available via the ESO Science Archive Facility.

%%%%%%%%%%%%%%%%%%%%%%%%%%%%%%%%%%%%%%%%%%%%%%%%%%

%%%%%%%%%%%%%%%%%%%% REFERENCES %%%%%%%%%%%%%%%%%%

% The best way to enter references is to use BibTeX:

\bibliographystyle{mnras}
\bibliography{Mypaperlib.bib} 

\begin{thebibliography}{}
\makeatletter
\relax
\def\mn@urlcharsother{\let\do\@makeother \do\$\do\&\do\#\do\^\do\_\do\%\do\~}
\def\mn@doi{\begingroup\mn@urlcharsother \@ifnextchar [ {\mn@doi@}
  {\mn@doi@[]}}
\def\mn@doi@[#1]#2{\def\@tempa{#1}\ifx\@tempa\@empty \href
  {http://dx.doi.org/#2} {doi:#2}\else \href {http://dx.doi.org/#2} {#1}\fi
  \endgroup}
\def\mn@eprint#1#2{\mn@eprint@#1:#2::\@nil}
\def\mn@eprint@arXiv#1{\href {http://arxiv.org/abs/#1} {{\tt arXiv:#1}}}
\def\mn@eprint@dblp#1{\href {http://dblp.uni-trier.de/rec/bibtex/#1.xml}
  {dblp:#1}}
\def\mn@eprint@#1:#2:#3:#4\@nil{\def\@tempa {#1}\def\@tempb {#2}\def\@tempc
  {#3}\ifx \@tempc \@empty \let \@tempc \@tempb \let \@tempb \@tempa \fi \ifx
  \@tempb \@empty \def\@tempb {arXiv}\fi \@ifundefined
  {mn@eprint@\@tempb}{\@tempb:\@tempc}{\expandafter \expandafter \csname
  mn@eprint@\@tempb\endcsname \expandafter{\@tempc}}}

\bibitem[\protect\citeauthoryear{{Abbott} et~al.,}{{Abbott}
  et~al.}{2018}]{Abbott18}
{Abbott} T.~M.~C.,  et~al., 2018, \mn@doi [\apjs] {10.3847/1538-4365/aae9f0},
  \href {https://ui.adsabs.harvard.edu/abs/2018ApJS..239...18A} {239, 18}

\bibitem[\protect\citeauthoryear{{Akaike}}{{Akaike}}{1974}]{Akaike74}
{Akaike} H.,  1974, IEEE Transactions on Automatic Control, \href
  {https://ui.adsabs.harvard.edu/abs/1974ITAC...19..716A} {19, 716}

\bibitem[\protect\citeauthoryear{{Astropy Collaboration} et~al.,}{{Astropy
  Collaboration} et~al.}{2013}]{Astropy-Collaboration13}
{Astropy Collaboration} et~al., 2013, \mn@doi [\aap]
  {10.1051/0004-6361/201322068}, \href
  {http://adsabs.harvard.edu/abs/2013A%26A...558A..33A} {558, A33}

\bibitem[\protect\citeauthoryear{{Bacon} et~al.,}{{Bacon}
  et~al.}{2010}]{Bacon10}
{Bacon} R.,  et~al., 2010, in Ground-based and Airborne Instrumentation for
  Astronomy III. p. 773508, \mn@doi{10.1117/12.856027}

\bibitem[\protect\citeauthoryear{{Bacon} et~al.,}{{Bacon}
  et~al.}{2017}]{Bacon17}
{Bacon} R.,  et~al., 2017, \mn@doi [\aap] {10.1051/0004-6361/201730833}, \href
  {https://ui.adsabs.harvard.edu/abs/2017A%26A...608A...1B} {608, A1}

\bibitem[\protect\citeauthoryear{{Balogh} et~al.,}{{Balogh}
  et~al.}{2004}]{Balogh04}
{Balogh} M.,  et~al., 2004, \mn@doi [\mnras]
  {10.1111/j.1365-2966.2004.07453.x}, \href
  {http://adsabs.harvard.edu/abs/2004MNRAS.348.1355B} {348, 1355}

\bibitem[\protect\citeauthoryear{{Balogh} et~al.,}{{Balogh}
  et~al.}{2014}]{Balogh14}
{Balogh} M.~L.,  et~al., 2014, \mn@doi [\mnras] {10.1093/mnras/stu1332}, \href
  {http://adsabs.harvard.edu/abs/2014MNRAS.443.2679B} {443, 2679}

\bibitem[\protect\citeauthoryear{{Bergeron}}{{Bergeron}}{1986}]{Bergeron86}
{Bergeron} J.,  1986, \aap, \href
  {https://ui.adsabs.harvard.edu/abs/1986A%26A...155L...8B} {155, L8}

\bibitem[\protect\citeauthoryear{{Berlind} et~al.,}{{Berlind}
  et~al.}{2006}]{Berlind06}
{Berlind} A.~A.,  et~al., 2006, \mn@doi [\apjs] {10.1086/508170}, \href
  {http://adsabs.harvard.edu/abs/2006ApJS..167....1B} {167, 1}

\bibitem[\protect\citeauthoryear{{Bertin} \& {Arnouts}}{{Bertin} \&
  {Arnouts}}{1996}]{Bertin96}
{Bertin} E.,  {Arnouts} S.,  1996, \mn@doi [\aaps] {10.1051/aas:1996164}, \href
  {http://adsabs.harvard.edu/abs/1996A%26AS..117..393B} {117, 393}

\bibitem[\protect\citeauthoryear{{Bielby}, {Crighton}, {Fumagalli}, {Morris},
  {Stott}, {Tejos}  \& {Cantalupo}}{{Bielby} et~al.}{2017}]{Bielby17a}
{Bielby} R.,  {Crighton} N.~H.~M.,  {Fumagalli} M.,  {Morris} S.~L.,  {Stott}
  J.~P.,  {Tejos} N.,   {Cantalupo} S.,  2017, \mn@doi [\mnras]
  {10.1093/mnras/stx528}, \href
  {https://ui.adsabs.harvard.edu/abs/2017MNRAS.468.1373B} {468, 1373}

\bibitem[\protect\citeauthoryear{{Bordoloi} et~al.,}{{Bordoloi}
  et~al.}{2011}]{Bordoloi11}
{Bordoloi} R.,  et~al., 2011, \mn@doi [\apj] {10.1088/0004-637X/743/1/10},
  \href {https://ui.adsabs.harvard.edu/abs/2011ApJ...743...10B} {743, 10}

\bibitem[\protect\citeauthoryear{{Bordoloi}, {Lilly}, {Kacprzak}  \&
  {Churchill}}{{Bordoloi} et~al.}{2014a}]{Bordoloi14a}
{Bordoloi} R.,  {Lilly} S.~J.,  {Kacprzak} G.~G.,   {Churchill} C.~W.,  2014a,
  \mn@doi [\apj] {10.1088/0004-637X/784/2/108}, \href
  {https://ui.adsabs.harvard.edu/abs/2014ApJ...784..108B} {784, 108}

\bibitem[\protect\citeauthoryear{{Bordoloi} et~al.,}{{Bordoloi}
  et~al.}{2014b}]{Bordoloi14}
{Bordoloi} R.,  et~al., 2014b, \mn@doi [\apj] {10.1088/0004-637X/796/2/136},
  \href {https://ui.adsabs.harvard.edu/abs/2014ApJ...796..136B} {796, 136}

\bibitem[\protect\citeauthoryear{{Borisova} et~al.,}{{Borisova}
  et~al.}{2016}]{Borisova16}
{Borisova} E.,  et~al., 2016, \mn@doi [\apj] {10.3847/0004-637X/831/1/39},
  \href {http://adsabs.harvard.edu/abs/2016ApJ...831...39B} {831, 39}

\bibitem[\protect\citeauthoryear{{Boselli} \& {Gavazzi}}{{Boselli} \&
  {Gavazzi}}{2006}]{Boselli06}
{Boselli} A.,  {Gavazzi} G.,  2006, \mn@doi [\pasp] {10.1086/500691}, \href
  {http://adsabs.harvard.edu/abs/2006PASP..118..517B} {118, 517}

\bibitem[\protect\citeauthoryear{{Bouch{\'e}} et~al.,}{{Bouch{\'e}}
  et~al.}{2010}]{Bouche10}
{Bouch{\'e}} N.,  et~al., 2010, \mn@doi [\apj] {10.1088/0004-637X/718/2/1001},
  \href {http://adsabs.harvard.edu/abs/2010ApJ...718.1001B} {718, 1001}

\bibitem[\protect\citeauthoryear{{Brammer}, {van Dokkum}  \& {Coppi}}{{Brammer}
  et~al.}{2008}]{Brammer08}
{Brammer} G.~B.,  {van Dokkum} P.~G.,   {Coppi} P.,  2008, \mn@doi [\apj]
  {10.1086/591786}, \href {http://adsabs.harvard.edu/abs/2008ApJ...686.1503B}
  {686, 1503}

\bibitem[\protect\citeauthoryear{{Bruzual} \& {Charlot}}{{Bruzual} \&
  {Charlot}}{2003}]{Bruzual03}
{Bruzual} G.,  {Charlot} S.,  2003, \mn@doi [\mnras]
  {10.1046/j.1365-8711.2003.06897.x}, \href
  {http://adsabs.harvard.edu/abs/2003MNRAS.344.1000B} {344, 1000}

\bibitem[\protect\citeauthoryear{{Buchner} et~al.,}{{Buchner}
  et~al.}{2014}]{Buchner14}
{Buchner} J.,  et~al., 2014, \mn@doi [\aap] {10.1051/0004-6361/201322971},
  \href {http://adsabs.harvard.edu/abs/2014A%26A...564A.125B} {564, A125}

\bibitem[\protect\citeauthoryear{{Byler}, {Dalcanton}, {Conroy}, {Johnson},
  {Levesque}  \& {Berg}}{{Byler} et~al.}{2018}]{Byler18}
{Byler} N.,  {Dalcanton} J.~J.,  {Conroy} C.,  {Johnson} B.~D.,  {Levesque}
  E.~M.,   {Berg} D.~A.,  2018, \mn@doi [\apj] {10.3847/1538-4357/aacd50},
  \href {https://ui.adsabs.harvard.edu/abs/2018ApJ...863...14B} {863, 14}

\bibitem[\protect\citeauthoryear{{Calzetti}, {Armus}, {Bohlin}, {Kinney},
  {Koornneef}  \& {Storchi-Bergmann}}{{Calzetti} et~al.}{2000}]{calzetti00}
{Calzetti} D.,  {Armus} L.,  {Bohlin} R.~C.,  {Kinney} A.~L.,  {Koornneef} J.,
   {Storchi-Bergmann} T.,  2000, \mn@doi [\apj] {10.1086/308692}, \href
  {http://adsabs.harvard.edu/abs/2000ApJ...533..682C} {533, 682}

\bibitem[\protect\citeauthoryear{{Cantalupo} et~al.,}{{Cantalupo}
  et~al.}{2019}]{Cantalupo19}
{Cantalupo} S.,  et~al., 2019, \mn@doi [\mnras] {10.1093/mnras/sty3481}, \href
  {https://ui.adsabs.harvard.edu/abs/2019MNRAS.483.5188C} {483, 5188}

\bibitem[\protect\citeauthoryear{{Chabrier}}{{Chabrier}}{2003}]{Chabrier03}
{Chabrier} G.,  2003, \mn@doi [\pasp] {10.1086/376392}, \href
  {http://adsabs.harvard.edu/abs/2003PASP..115..763C} {115, 763}

\bibitem[\protect\citeauthoryear{{Chen} \& {Tinker}}{{Chen} \&
  {Tinker}}{2008}]{Chen08}
{Chen} H.-W.,  {Tinker} J.~L.,  2008, \mn@doi [\apj] {10.1086/591927}, \href
  {https://ui.adsabs.harvard.edu/abs/2008ApJ...687..745C} {687, 745}

\bibitem[\protect\citeauthoryear{{Chen}, {Helsby}, {Gauthier}, {Shectman},
  {Thompson}  \& {Tinker}}{{Chen} et~al.}{2010}]{Chen10}
{Chen} H.-W.,  {Helsby} J.~E.,  {Gauthier} J.-R.,  {Shectman} S.~A.,
  {Thompson} I.~B.,   {Tinker} J.~L.,  2010, \mn@doi [\apj]
  {10.1088/0004-637X/714/2/1521}, \href
  {https://ui.adsabs.harvard.edu/abs/2010ApJ...714.1521C} {714, 1521}

\bibitem[\protect\citeauthoryear{{Chen}, {Boettcher}, {Johnson}, {Zahedy},
  {Rudie}, {Cooksey}, {Rauch}  \& {Mulchaey}}{{Chen} et~al.}{2019}]{Chen19a}
{Chen} H.-W.,  {Boettcher} E.,  {Johnson} S.~D.,  {Zahedy} F.~S.,  {Rudie}
  G.~C.,  {Cooksey} K.~L.,  {Rauch} M.,   {Mulchaey} J.~S.,  2019, \mn@doi
  [\apjl] {10.3847/2041-8213/ab25ec}, \href
  {https://ui.adsabs.harvard.edu/abs/2019ApJ...878L..33C} {878, L33}

\bibitem[\protect\citeauthoryear{{Chiang}, {Overzier}  \& {Gebhardt}}{{Chiang}
  et~al.}{2013}]{Chiang13}
{Chiang} Y.-K.,  {Overzier} R.,   {Gebhardt} K.,  2013, \mn@doi [\apj]
  {10.1088/0004-637X/779/2/127}, \href
  {https://ui.adsabs.harvard.edu/abs/2013ApJ...779..127C} {779, 127}

\bibitem[\protect\citeauthoryear{{Colbert}, {Scarlata}, {Teplitz}, {Francis},
  {Palunas}, {Williger}  \& {Woodgate}}{{Colbert} et~al.}{2011}]{Colbert11}
{Colbert} J.~W.,  {Scarlata} C.,  {Teplitz} H.,  {Francis} P.,  {Palunas} P.,
  {Williger} G.~M.,   {Woodgate} B.,  2011, \mn@doi [\apj]
  {10.1088/0004-637X/728/1/59}, \href
  {https://ui.adsabs.harvard.edu/abs/2011ApJ...728...59C} {728, 59}

\bibitem[\protect\citeauthoryear{{Conroy}, {Gunn}  \& {White}}{{Conroy}
  et~al.}{2009}]{Conroy09}
{Conroy} C.,  {Gunn} J.~E.,   {White} M.,  2009, \mn@doi [\apj]
  {10.1088/0004-637X/699/1/486}, \href
  {http://adsabs.harvard.edu/abs/2009ApJ...699..486C} {699, 486}

\bibitem[\protect\citeauthoryear{{Crighton} et~al.,}{{Crighton}
  et~al.}{2011}]{Crighton11}
{Crighton} N.~H.~M.,  et~al., 2011, \mn@doi [\mnras]
  {10.1111/j.1365-2966.2011.17247.x}, \href
  {https://ui.adsabs.harvard.edu/abs/2011MNRAS.414...28C} {414, 28}

\bibitem[\protect\citeauthoryear{{Cucciati} et~al.,}{{Cucciati}
  et~al.}{2014}]{Cucciati14}
{Cucciati} O.,  et~al., 2014, \mn@doi [\aap] {10.1051/0004-6361/201423409},
  \href {http://adsabs.harvard.edu/abs/2014A%26A...565A..67C} {565, A67}

\bibitem[\protect\citeauthoryear{{Cupani} et~al.,}{{Cupani}
  et~al.}{2016}]{Cupani16}
{Cupani} G.,  et~al., 2016, in Software and Cyberinfrastructure for Astronomy
  IV. p. 99131T, \mn@doi{10.1117/12.2231379}

\bibitem[\protect\citeauthoryear{{D'Odorico}, {Petitjean}  \&
  {Cristiani}}{{D'Odorico} et~al.}{2002}]{DOdorico02}
{D'Odorico} V.,  {Petitjean} P.,   {Cristiani} S.,  2002, \mn@doi [\aap]
  {10.1051/0004-6361:20020737}, \href
  {https://ui.adsabs.harvard.edu/abs/2002A%26A...390...13D} {390, 13}

\bibitem[\protect\citeauthoryear{{Dav{\'e}}, {Finlator}  \&
  {Oppenheimer}}{{Dav{\'e}} et~al.}{2012}]{Dave12}
{Dav{\'e}} R.,  {Finlator} K.,   {Oppenheimer} B.~D.,  2012, \mn@doi [\mnras]
  {10.1111/j.1365-2966.2011.20148.x}, \href
  {https://ui.adsabs.harvard.edu/abs/2012MNRAS.421...98D} {421, 98}

\bibitem[\protect\citeauthoryear{{Dekel} \& {Birnboim}}{{Dekel} \&
  {Birnboim}}{2006}]{Dekel06}
{Dekel} A.,  {Birnboim} Y.,  2006, \mn@doi [\mnras]
  {10.1111/j.1365-2966.2006.10145.x}, \href
  {http://adsabs.harvard.edu/abs/2006MNRAS.368....2D} {368, 2}

\bibitem[\protect\citeauthoryear{{Dekel} \& {Silk}}{{Dekel} \&
  {Silk}}{1986}]{Dekel86}
{Dekel} A.,  {Silk} J.,  1986, \mn@doi [\apj] {10.1086/164050}, \href
  {https://ui.adsabs.harvard.edu/abs/1986ApJ...303...39D} {303, 39}

\bibitem[\protect\citeauthoryear{{Dekker}, {D'Odorico}, {Kaufer}, {Delabre}  \&
  {Kotzlowski}}{{Dekker} et~al.}{2000}]{Dekker00}
{Dekker} H.,  {D'Odorico} S.,  {Kaufer} A.,  {Delabre} B.,   {Kotzlowski} H.,
  2000, in {Iye} M.,  {Moorwood} A.~F.,  eds,  \procspie Vol. 4008, Optical and
  IR Telescope Instrumentation and Detectors. pp 534--545,
  \mn@doi{10.1117/12.395512}

\bibitem[\protect\citeauthoryear{{Diener} et~al.,}{{Diener}
  et~al.}{2013}]{Diener13}
{Diener} C.,  et~al., 2013, \mn@doi [\apj] {10.1088/0004-637X/765/2/109}, \href
  {http://adsabs.harvard.edu/abs/2013ApJ...765..109D} {765, 109}

\bibitem[\protect\citeauthoryear{{Dressler}}{{Dressler}}{1980}]{Dressler80}
{Dressler} A.,  1980, \mn@doi [\apj] {10.1086/157753}, \href
  {http://adsabs.harvard.edu/abs/1980ApJ...236..351D} {236, 351}

\bibitem[\protect\citeauthoryear{{Driver} et~al.,}{{Driver}
  et~al.}{2011}]{Driver11}
{Driver} S.~P.,  et~al., 2011, \mn@doi [\mnras]
  {10.1111/j.1365-2966.2010.18188.x}, \href
  {https://ui.adsabs.harvard.edu/abs/2011MNRAS.413..971D} {413, 971}

\bibitem[\protect\citeauthoryear{{Fazio} et~al.,}{{Fazio}
  et~al.}{2004}]{Fazio04}
{Fazio} G.~G.,  et~al., 2004, \mn@doi [\apjs] {10.1086/422843}, \href
  {http://adsabs.harvard.edu/abs/2004ApJS..154...10F} {154, 10}

\bibitem[\protect\citeauthoryear{{Feroz} \& {Hobson}}{{Feroz} \&
  {Hobson}}{2008}]{Feroz08}
{Feroz} F.,  {Hobson} M.~P.,  2008, \mn@doi [\mnras]
  {10.1111/j.1365-2966.2007.12353.x}, \href
  {http://adsabs.harvard.edu/abs/2008MNRAS.384..449F} {384, 449}

\bibitem[\protect\citeauthoryear{{Feroz}, {Hobson}, {Cameron}  \&
  {Pettitt}}{{Feroz} et~al.}{2013}]{Feroz13}
{Feroz} F.,  {Hobson} M.~P.,  {Cameron} E.,   {Pettitt} A.~N.,  2013, preprint,
  \href {http://adsabs.harvard.edu/abs/2013arXiv1306.2144F} {} (\mn@eprint
  {arXiv} {1306.2144})

\bibitem[\protect\citeauthoryear{{Finn} et~al.,}{{Finn} et~al.}{2016}]{Finn16}
{Finn} C.~W.,  et~al., 2016, \mn@doi [\mnras] {10.1093/mnras/stw918}, \href
  {https://ui.adsabs.harvard.edu/abs/2016MNRAS.460..590F} {460, 590}

\bibitem[\protect\citeauthoryear{{Fossati} et~al.,}{{Fossati}
  et~al.}{2017}]{Fossati17}
{Fossati} M.,  et~al., 2017, \mn@doi [\apj] {10.3847/1538-4357/835/2/153},
  \href {http://adsabs.harvard.edu/abs/2017ApJ...835..153F} {835, 153}

\bibitem[\protect\citeauthoryear{{Fossati} et~al.,}{{Fossati}
  et~al.}{2018}]{Fossati18}
{Fossati} M.,  et~al., 2018, \mn@doi [\aap] {10.1051/0004-6361/201732373},
  \href {http://adsabs.harvard.edu/abs/2018A%26A...614A..57F} {614, A57}

\bibitem[\protect\citeauthoryear{{Fossati}, {Fumagalli}, {Gavazzi},
  {Consolandi}, {Boselli}, {Yagi}, {Sun}  \& {Wilman}}{{Fossati}
  et~al.}{2019}]{Fossati19}
{Fossati} M.,  {Fumagalli} M.,  {Gavazzi} G.,  {Consolandi} G.,  {Boselli} A.,
  {Yagi} M.,  {Sun} M.,   {Wilman} D.~J.,  2019, \mn@doi [\mnras]
  {10.1093/mnras/stz136}, \href
  {http://adsabs.harvard.edu/abs/2019MNRAS.484.2212F} {484, 2212}

\bibitem[\protect\citeauthoryear{{Francis} \& {Hewett}}{{Francis} \&
  {Hewett}}{1993}]{Francis93}
{Francis} P.~J.,  {Hewett} P.~C.,  1993, \mn@doi [\aj] {10.1086/116542}, \href
  {https://ui.adsabs.harvard.edu/abs/1993AJ....105.1633F} {105, 1633}

\bibitem[\protect\citeauthoryear{{Fraternali} \& {Binney}}{{Fraternali} \&
  {Binney}}{2008}]{Fraternali08}
{Fraternali} F.,  {Binney} J.~J.,  2008, \mn@doi [\mnras]
  {10.1111/j.1365-2966.2008.13071.x}, \href
  {https://ui.adsabs.harvard.edu/abs/2008MNRAS.386..935F} {386, 935}

\bibitem[\protect\citeauthoryear{{Fumagalli}, {Cantalupo}, {Dekel}, {Morris},
  {O'Meara}, {Prochaska}  \& {Theuns}}{{Fumagalli} et~al.}{2016}]{Fumagalli16}
{Fumagalli} M.,  {Cantalupo} S.,  {Dekel} A.,  {Morris} S.~L.,  {O'Meara}
  J.~M.,  {Prochaska} J.~X.,   {Theuns} T.,  2016, \mn@doi [\mnras]
  {10.1093/mnras/stw1782}, \href
  {https://ui.adsabs.harvard.edu/abs/2016MNRAS.462.1978F} {462, 1978}

\bibitem[\protect\citeauthoryear{{Fumagalli}, {Haardt}, {Theuns}, {Morris},
  {Cantalupo}, {Madau}  \& {Fossati}}{{Fumagalli} et~al.}{2017a}]{Fumagalli17}
{Fumagalli} M.,  {Haardt} F.,  {Theuns} T.,  {Morris} S.~L.,  {Cantalupo} S.,
  {Madau} P.,   {Fossati} M.,  2017a, \mn@doi [\mnras] {10.1093/mnras/stx398},
  \href {http://adsabs.harvard.edu/abs/2017MNRAS.467.4802F} {467, 4802}

\bibitem[\protect\citeauthoryear{{Fumagalli} et~al.,}{{Fumagalli}
  et~al.}{2017b}]{Fumagalli17a}
{Fumagalli} M.,  et~al., 2017b, \mn@doi [\mnras] {10.1093/mnras/stx1896}, \href
  {https://ui.adsabs.harvard.edu/abs/2017MNRAS.471.3686F} {471, 3686}

\bibitem[\protect\citeauthoryear{{Gaia Collaboration} et~al.,}{{Gaia
  Collaboration} et~al.}{2018}]{Gaia18}
{Gaia Collaboration} et~al., 2018, \mn@doi [\aap]
  {10.1051/0004-6361/201833051}, \href
  {https://ui.adsabs.harvard.edu/abs/2018A%26A...616A...1G} {616, A1}

\bibitem[\protect\citeauthoryear{{Galametz} et~al.,}{{Galametz}
  et~al.}{2018}]{Galametz18}
{Galametz} A.,  et~al., 2018, \mn@doi [\mnras] {10.1093/mnras/sty095}, \href
  {http://adsabs.harvard.edu/abs/2018MNRAS.475.4148G} {475, 4148}

\bibitem[\protect\citeauthoryear{{Gallazzi} \& {Bell}}{{Gallazzi} \&
  {Bell}}{2009}]{Gallazzi09}
{Gallazzi} A.,  {Bell} E.~F.,  2009, \mn@doi [\apjs]
  {10.1088/0067-0049/185/2/253}, \href
  {http://adsabs.harvard.edu/abs/2009ApJS..185..253G} {185, 253}

\bibitem[\protect\citeauthoryear{{Gauthier}}{{Gauthier}}{2013}]{Gauthier13}
{Gauthier} J.-R.,  2013, \mn@doi [\mnras] {10.1093/mnras/stt565}, \href
  {https://ui.adsabs.harvard.edu/abs/2013MNRAS.432.1444G} {432, 1444}

\bibitem[\protect\citeauthoryear{{Giovanelli} \& {Haynes}}{{Giovanelli} \&
  {Haynes}}{1985}]{Giovanelli85}
{Giovanelli} R.,  {Haynes} M.~P.,  1985, \mn@doi [\apj] {10.1086/163170}, \href
  {http://adsabs.harvard.edu/abs/1985ApJ...292..404G} {292, 404}

\bibitem[\protect\citeauthoryear{{Gunn} \& {Gott}}{{Gunn} \&
  {Gott}}{1972}]{Gunn72}
{Gunn} J.~E.,  {Gott} III J.~R.,  1972, \mn@doi [\apj] {10.1086/151605}, \href
  {http://adsabs.harvard.edu/abs/1972ApJ...176....1G} {176, 1}

\bibitem[\protect\citeauthoryear{{Handley}, {Hobson}  \& {Lasenby}}{{Handley}
  et~al.}{2015}]{Handley15}
{Handley} W.~J.,  {Hobson} M.~P.,   {Lasenby} A.~N.,  2015, \mn@doi [\mnras]
  {10.1093/mnras/stv1911}, \href
  {https://ui.adsabs.harvard.edu/abs/2015MNRAS.453.4384H} {453, 4384}

\bibitem[\protect\citeauthoryear{{Henriques}, {White}, {Thomas}, {Angulo},
  {Guo}, {Lemson}, {Springel}  \& {Overzier}}{{Henriques}
  et~al.}{2015}]{Henriques15}
{Henriques} B.~M.~B.,  {White} S.~D.~M.,  {Thomas} P.~A.,  {Angulo} R.,  {Guo}
  Q.,  {Lemson} G.,  {Springel} V.,   {Overzier} R.,  2015, \mn@doi [\mnras]
  {10.1093/mnras/stv705}, \href
  {http://adsabs.harvard.edu/abs/2015MNRAS.451.2663H} {451, 2663}

\bibitem[\protect\citeauthoryear{{Hinton}, {Davis}, {Lidman}, {Glazebrook}  \&
  {Lewis}}{{Hinton} et~al.}{2016}]{Hinton16}
{Hinton} S.~R.,  {Davis} T.~M.,  {Lidman} C.,  {Glazebrook} K.,   {Lewis}
  G.~F.,  2016, \mn@doi [Astronomy and Computing]
  {10.1016/j.ascom.2016.03.001}, \href
  {https://ui.adsabs.harvard.edu/abs/2016A%26C....15...61H} {15, 61}

\bibitem[\protect\citeauthoryear{{Hoyle} et~al.,}{{Hoyle}
  et~al.}{2018}]{Hoyle18}
{Hoyle} B.,  et~al., 2018, \mn@doi [\mnras] {10.1093/mnras/sty957}, \href
  {https://ui.adsabs.harvard.edu/abs/2018MNRAS.478..592H} {478, 592}

\bibitem[\protect\citeauthoryear{{Huchra} \& {Geller}}{{Huchra} \&
  {Geller}}{1982}]{Huchra82}
{Huchra} J.~P.,  {Geller} M.~J.,  1982, \mn@doi [\apj] {10.1086/160000}, \href
  {https://ui.adsabs.harvard.edu/abs/1982ApJ...257..423H} {257, 423}

\bibitem[\protect\citeauthoryear{{Kacprzak}, {Churchill}, {Steidel}  \&
  {Murphy}}{{Kacprzak} et~al.}{2008}]{Kacprzak08}
{Kacprzak} G.~G.,  {Churchill} C.~W.,  {Steidel} C.~C.,   {Murphy} M.~T.,
  2008, \mn@doi [\aj] {10.1088/0004-6256/135/3/922}, \href
  {https://ui.adsabs.harvard.edu/abs/2008AJ....135..922K} {135, 922}

\bibitem[\protect\citeauthoryear{{Kacprzak}, {Murphy}  \&
  {Churchill}}{{Kacprzak} et~al.}{2010}]{Kacprzak10}
{Kacprzak} G.~G.,  {Murphy} M.~T.,   {Churchill} C.~W.,  2010, \mn@doi [\mnras]
  {10.1111/j.1365-2966.2010.16667.x}, \href
  {https://ui.adsabs.harvard.edu/abs/2010MNRAS.406..445K} {406, 445}

\bibitem[\protect\citeauthoryear{{Kauffmann}, {Nelson}, {M{\'e}nard}  \&
  {Zhu}}{{Kauffmann} et~al.}{2017}]{Kauffmann17}
{Kauffmann} G.,  {Nelson} D.,  {M{\'e}nard} B.,   {Zhu} G.,  2017, \mn@doi
  [\mnras] {10.1093/mnras/stx639}, \href
  {https://ui.adsabs.harvard.edu/abs/2017MNRAS.468.3737K} {468, 3737}

\bibitem[\protect\citeauthoryear{{Kere{\v s}}, {Katz}, {Weinberg}  \&
  {Dav{\'e}}}{{Kere{\v s}} et~al.}{2005}]{Keres05}
{Kere{\v s}} D.,  {Katz} N.,  {Weinberg} D.,   {Dav{\'e}} R.,  2005, \mn@doi
  [\mnras] {10.1111/j.1365-2966.2005.09451.x}, \href
  {https://ui.adsabs.harvard.edu/abs/2005MNRAS.363....2K} {363, 2}

\bibitem[\protect\citeauthoryear{{Klitsch}, {P{\'e}roux}, {Zwaan}, {Smail},
  {Oteo}, {Biggs}, {Popping}  \& {Swinbank}}{{Klitsch}
  et~al.}{2018}]{Klitsch18}
{Klitsch} A.,  {P{\'e}roux} C.,  {Zwaan} M.~A.,  {Smail} I.,  {Oteo} I.,
  {Biggs} A.~D.,  {Popping} G.,   {Swinbank} A.~M.,  2018, \mn@doi [\mnras]
  {10.1093/mnras/stx3184}, \href
  {https://ui.adsabs.harvard.edu/abs/2018MNRAS.475..492K} {475, 492}

\bibitem[\protect\citeauthoryear{{Knobel} et~al.,}{{Knobel}
  et~al.}{2009}]{Knobel09}
{Knobel} C.,  et~al., 2009, \mn@doi [\apj] {10.1088/0004-637X/697/2/1842},
  \href {https://ui.adsabs.harvard.edu/abs/2009ApJ...697.1842K} {697, 1842}

\bibitem[\protect\citeauthoryear{{Knobel} et~al.,}{{Knobel}
  et~al.}{2012}]{Knobel12}
{Knobel} C.,  et~al., 2012, \mn@doi [\apj] {10.1088/0004-637X/753/2/121}, \href
  {http://adsabs.harvard.edu/abs/2012ApJ...753..121K} {753, 121}

\bibitem[\protect\citeauthoryear{{Kova{\v c}} et~al.,}{{Kova{\v c}}
  et~al.}{2010}]{Kovac10}
{Kova{\v c}} K.,  et~al., 2010, \mn@doi [\apj] {10.1088/0004-637X/708/1/505},
  \href {http://adsabs.harvard.edu/abs/2010ApJ...708..505K} {708, 505}

\bibitem[\protect\citeauthoryear{{Kron}}{{Kron}}{1980}]{Kron80}
{Kron} R.~G.,  1980, \mn@doi [\apjs] {10.1086/190669}, \href
  {https://ui.adsabs.harvard.edu/abs/1980ApJS...43..305K} {43, 305}

\bibitem[\protect\citeauthoryear{{Lilly}, {Carollo}, {Pipino}, {Renzini}  \&
  {Peng}}{{Lilly} et~al.}{2013}]{Lilly13}
{Lilly} S.~J.,  {Carollo} C.~M.,  {Pipino} A.,  {Renzini} A.,   {Peng} Y.,
  2013, \mn@doi [\apj] {10.1088/0004-637X/772/2/119}, \href
  {http://adsabs.harvard.edu/abs/2013ApJ...772..119L} {772, 119}

\bibitem[\protect\citeauthoryear{{Lofthouse}, {Fumagalli}, {Fossati},
  {O'Meara}, {Murphy}  \& et al.}{{Lofthouse} et~al.}{2019}]{Lofthouse19}
{Lofthouse} E.,  {Fumagalli} M.,  {Fossati} M.,  {O'Meara} J.,  {Murphy} M.,
  et al. 2019, \mnras, p. submitted

\bibitem[\protect\citeauthoryear{{Lusso} et~al.,}{{Lusso}
  et~al.}{2019}]{Lusso19}
{Lusso} E.,  et~al., 2019, \mn@doi [\mnras] {10.1093/mnrasl/slz032}, \href
  {http://adsabs.harvard.edu/abs/2019MNRAS.485L..62L} {485, L62}

\bibitem[\protect\citeauthoryear{{McGee}, {Bower}  \& {Balogh}}{{McGee}
  et~al.}{2014}]{Mcgee14}
{McGee} S.~L.,  {Bower} R.~G.,   {Balogh} M.~L.,  2014, \mn@doi [\mnras]
  {10.1093/mnrasl/slu066}, \href
  {http://adsabs.harvard.edu/abs/2014MNRAS.442L.105M} {442, L105}

\bibitem[\protect\citeauthoryear{{Mendel}, {Simard}, {Palmer}, {Ellison}  \&
  {Patton}}{{Mendel} et~al.}{2014}]{Mendel14}
{Mendel} J.~T.,  {Simard} L.,  {Palmer} M.,  {Ellison} S.~L.,   {Patton} D.~R.,
   2014, \mn@doi [\apjs] {10.1088/0067-0049/210/1/3}, \href
  {http://adsabs.harvard.edu/abs/2014ApJS..210....3M} {210, 3}

\bibitem[\protect\citeauthoryear{{Merlin} et~al.,}{{Merlin}
  et~al.}{2015}]{Merlin15}
{Merlin} E.,  et~al., 2015, \mn@doi [\aap] {10.1051/0004-6361/201526471}, \href
  {http://adsabs.harvard.edu/abs/2015A%26A...582A..15M} {582, A15}

\bibitem[\protect\citeauthoryear{{Mihos}, {Keating}, {Holley-Bockelmann},
  {Pisano}  \& {Kassim}}{{Mihos} et~al.}{2012}]{Mihos12}
{Mihos} J.~C.,  {Keating} K.~M.,  {Holley-Bockelmann} K.,  {Pisano} D.~J.,
  {Kassim} N.~E.,  2012, \mn@doi [\apj] {10.1088/0004-637X/761/2/186}, \href
  {https://ui.adsabs.harvard.edu/abs/2012ApJ...761..186M} {761, 186}

\bibitem[\protect\citeauthoryear{{Momcheva} et~al.,}{{Momcheva}
  et~al.}{2016}]{Momcheva16}
{Momcheva} I.~G.,  et~al., 2016, \mn@doi [\apjs] {10.3847/0067-0049/225/2/27},
  \href {http://adsabs.harvard.edu/abs/2016ApJS..225...27M} {225, 27}

\bibitem[\protect\citeauthoryear{{Morris} \& {van den Bergh}}{{Morris} \& {van
  den Bergh}}{1994}]{Morris94}
{Morris} S.~L.,  {van den Bergh} S.,  1994, \mn@doi [\apj] {10.1086/174176},
  \href {https://ui.adsabs.harvard.edu/abs/1994ApJ...427..696M} {427, 696}

\bibitem[\protect\citeauthoryear{{Morrissey} et~al.,}{{Morrissey}
  et~al.}{2018}]{Morrissey18}
{Morrissey} P.,  et~al., 2018, \mn@doi [\apj] {10.3847/1538-4357/aad597}, \href
  {https://ui.adsabs.harvard.edu/abs/2018ApJ...864...93M} {864, 93}

\bibitem[\protect\citeauthoryear{{Moster}, {Somerville}, {Maulbetsch}, {van den
  Bosch}, {Macci{\`o}}, {Naab}  \& {Oser}}{{Moster} et~al.}{2010}]{Moster10}
{Moster} B.~P.,  {Somerville} R.~S.,  {Maulbetsch} C.,  {van den Bosch} F.~C.,
  {Macci{\`o}} A.~V.,  {Naab} T.,   {Oser} L.,  2010, \mn@doi [\apj]
  {10.1088/0004-637X/710/2/903}, \href
  {http://adsabs.harvard.edu/abs/2010ApJ...710..903M} {710, 903}

\bibitem[\protect\citeauthoryear{{Muzzin} et~al.,}{{Muzzin}
  et~al.}{2013}]{Muzzin13}
{Muzzin} A.,  et~al., 2013, \mn@doi [\apj] {10.1088/0004-637X/777/1/18}, \href
  {http://adsabs.harvard.edu/abs/2013ApJ...777...18M} {777, 18}

\bibitem[\protect\citeauthoryear{{Nestor}, {Johnson}, {Wild}, {M{\'e}nard},
  {Turnshek}, {Rao}  \& {Pettini}}{{Nestor} et~al.}{2011}]{Nestor11}
{Nestor} D.~B.,  {Johnson} B.~D.,  {Wild} V.,  {M{\'e}nard} B.,  {Turnshek}
  D.~A.,  {Rao} S.,   {Pettini} M.,  2011, \mn@doi [\mnras]
  {10.1111/j.1365-2966.2010.17865.x}, \href
  {https://ui.adsabs.harvard.edu/abs/2011MNRAS.412.1559N} {412, 1559}

\bibitem[\protect\citeauthoryear{{Nielsen}, {Churchill}, {Kacprzak}, {Murphy}
  \& {Evans}}{{Nielsen} et~al.}{2015}]{Nielsen15}
{Nielsen} N.~M.,  {Churchill} C.~W.,  {Kacprzak} G.~G.,  {Murphy} M.~T.,
  {Evans} J.~L.,  2015, \mn@doi [\apj] {10.1088/0004-637X/812/1/83}, \href
  {https://ui.adsabs.harvard.edu/abs/2015ApJ...812...83N} {812, 83}

\bibitem[\protect\citeauthoryear{{Nielsen}, {Kacprzak}, {Pointon}, {Churchill}
  \& {Murphy}}{{Nielsen} et~al.}{2018}]{Nielsen18}
{Nielsen} N.~M.,  {Kacprzak} G.~G.,  {Pointon} S.~K.,  {Churchill} C.~W.,
  {Murphy} M.~T.,  2018, \mn@doi [\apj] {10.3847/1538-4357/aaedbd}, \href
  {https://ui.adsabs.harvard.edu/abs/2018ApJ...869..153N} {869, 153}

\bibitem[\protect\citeauthoryear{{Noble} et~al.,}{{Noble}
  et~al.}{2017}]{Noble17}
{Noble} A.~G.,  et~al., 2017, \mn@doi [\apjl] {10.3847/2041-8213/aa77f3}, \href
  {https://ui.adsabs.harvard.edu/abs/2017ApJ...842L..21N} {842, L21}

\bibitem[\protect\citeauthoryear{{Oemler}}{{Oemler}}{1974}]{Oemler74}
{Oemler} Jr. A.,  1974, \mn@doi [\apj] {10.1086/153216}, \href
  {http://adsabs.harvard.edu/abs/1974ApJ...194....1O} {194, 1}

\bibitem[\protect\citeauthoryear{{Oppenheimer}, {Dav{\'e}}, {Kere{\v s}},
  {Fardal}, {Katz}, {Kollmeier}  \& {Weinberg}}{{Oppenheimer}
  et~al.}{2010}]{Oppenheimer10}
{Oppenheimer} B.~D.,  {Dav{\'e}} R.,  {Kere{\v s}} D.,  {Fardal} M.,  {Katz}
  N.,  {Kollmeier} J.~A.,   {Weinberg} D.~H.,  2010, \mn@doi [\mnras]
  {10.1111/j.1365-2966.2010.16872.x}, \href
  {https://ui.adsabs.harvard.edu/abs/2010MNRAS.406.2325O} {406, 2325}

\bibitem[\protect\citeauthoryear{{Peng} et~al.,}{{Peng} et~al.}{2010}]{Peng10}
{Peng} Y.-j.,  et~al., 2010, \mn@doi [\apj] {10.1088/0004-637X/721/1/193},
  \href {http://adsabs.harvard.edu/abs/2010ApJ...721..193P} {721, 193}

\bibitem[\protect\citeauthoryear{{Perlmutter} et~al.,}{{Perlmutter}
  et~al.}{1999}]{Perlmutter99}
{Perlmutter} S.,  et~al., 1999, \mn@doi [\apj] {10.1086/307221}, \href
  {http://adsabs.harvard.edu/abs/1999ApJ...517..565P} {517, 565}

\bibitem[\protect\citeauthoryear{{P{\'e}roux} et~al.,}{{P{\'e}roux}
  et~al.}{2017}]{Peroux17}
{P{\'e}roux} C.,  et~al., 2017, \mn@doi [\mnras] {10.1093/mnras/stw2444}, \href
  {https://ui.adsabs.harvard.edu/abs/2017MNRAS.464.2053P} {464, 2053}

\bibitem[\protect\citeauthoryear{{Planck Collaboration} et~al.,}{{Planck
  Collaboration} et~al.}{2016}]{Planck16}
{Planck Collaboration} et~al., 2016, \mn@doi [\aap]
  {10.1051/0004-6361/201525830}, \href
  {https://ui.adsabs.harvard.edu/abs/2016A%26A...594A..13P} {594, A13}

\bibitem[\protect\citeauthoryear{{Prochaska}, {Weiner}, {Chen}, {Mulchaey}  \&
  {Cooksey}}{{Prochaska} et~al.}{2011}]{Prochaska11}
{Prochaska} J.~X.,  {Weiner} B.,  {Chen} H.-W.,  {Mulchaey} J.,   {Cooksey} K.,
   2011, \mn@doi [\apj] {10.1088/0004-637X/740/2/91}, \href
  {https://ui.adsabs.harvard.edu/abs/2011ApJ...740...91P} {740, 91}

\bibitem[\protect\citeauthoryear{{Rasmussen} et~al.,}{{Rasmussen}
  et~al.}{2012}]{Rasmussen12}
{Rasmussen} J.,  et~al., 2012, \mn@doi [\apj] {10.1088/0004-637X/747/1/31},
  \href {https://ui.adsabs.harvard.edu/abs/2012ApJ...747...31R} {747, 31}

\bibitem[\protect\citeauthoryear{{Rubin}, {Weiner}, {Koo}, {Martin},
  {Prochaska}, {Coil}  \& {Newman}}{{Rubin} et~al.}{2010}]{Rubin10}
{Rubin} K.~H.~R.,  {Weiner} B.~J.,  {Koo} D.~C.,  {Martin} C.~L.,  {Prochaska}
  J.~X.,  {Coil} A.~L.,   {Newman} J.~A.,  2010, \mn@doi [\apj]
  {10.1088/0004-637X/719/2/1503}, \href
  {https://ui.adsabs.harvard.edu/abs/2010ApJ...719.1503R} {719, 1503}

\bibitem[\protect\citeauthoryear{{Rubin}, {Diamond-Stanic}, {Coil}, {Crighton}
  \& {Moustakas}}{{Rubin} et~al.}{2018a}]{Rubin18}
{Rubin} K.~H.~R.,  {Diamond-Stanic} A.~M.,  {Coil} A.~L.,  {Crighton} N.~H.~M.,
    {Moustakas} J.,  2018a, \mn@doi [\apj] {10.3847/1538-4357/aa9792}, \href
  {https://ui.adsabs.harvard.edu/abs/2018ApJ...853...95R} {853, 95}

\bibitem[\protect\citeauthoryear{{Rubin}, {Diamond-Stanic}, {Coil}, {Crighton}
  \& {Stewart}}{{Rubin} et~al.}{2018b}]{Rubin18a}
{Rubin} K.~H.~R.,  {Diamond-Stanic} A.~M.,  {Coil} A.~L.,  {Crighton} N.~H.~M.,
    {Stewart} K.~R.,  2018b, \mn@doi [\apj] {10.3847/1538-4357/aad566}, \href
  {https://ui.adsabs.harvard.edu/abs/2018ApJ...868..142R} {868, 142}

\bibitem[\protect\citeauthoryear{{Rudie} et~al.,}{{Rudie}
  et~al.}{2012}]{Rudie12}
{Rudie} G.~C.,  et~al., 2012, \mn@doi [\apj] {10.1088/0004-637X/750/1/67},
  \href {https://ui.adsabs.harvard.edu/abs/2012ApJ...750...67R} {750, 67}

\bibitem[\protect\citeauthoryear{{Rudnick} et~al.,}{{Rudnick}
  et~al.}{2017}]{Rudnick17}
{Rudnick} G.,  et~al., 2017, \mn@doi [\apj] {10.3847/1538-4357/aa87b2}, \href
  {https://ui.adsabs.harvard.edu/abs/2017ApJ...849...27R} {849, 27}

\bibitem[\protect\citeauthoryear{{Schaye}, {Aguirre}, {Kim}, {Theuns}, {Rauch}
  \& {Sargent}}{{Schaye} et~al.}{2003}]{Schaye03}
{Schaye} J.,  {Aguirre} A.,  {Kim} T.-S.,  {Theuns} T.,  {Rauch} M.,
  {Sargent} W.~L.~W.,  2003, \mn@doi [\apj] {10.1086/378044}, \href
  {https://ui.adsabs.harvard.edu/abs/2003ApJ...596..768S} {596, 768}

\bibitem[\protect\citeauthoryear{{Schroetter} et~al.,}{{Schroetter}
  et~al.}{2016}]{Schroetter16}
{Schroetter} I.,  et~al., 2016, \mn@doi [\apj] {10.3847/1538-4357/833/1/39},
  \href {https://ui.adsabs.harvard.edu/abs/2016ApJ...833...39S} {833, 39}

\bibitem[\protect\citeauthoryear{{Scoville} et~al.,}{{Scoville}
  et~al.}{2007}]{Scoville07}
{Scoville} N.,  et~al., 2007, \mn@doi [\apjs] {10.1086/516751}, \href
  {http://adsabs.harvard.edu/abs/2007ApJS..172..150S} {172, 150}

\bibitem[\protect\citeauthoryear{{Sharma} \& {Theuns}}{{Sharma} \&
  {Theuns}}{2019}]{Sharma19}
{Sharma} M.,  {Theuns} T.,  2019, arXiv e-prints, \href
  {https://ui.adsabs.harvard.edu/abs/2019arXiv190610135S} {}

\bibitem[\protect\citeauthoryear{{Springel} et~al.,}{{Springel}
  et~al.}{2005}]{Springel05}
{Springel} V.,  et~al., 2005, \mn@doi [\nat] {10.1038/nature03597}, \href
  {https://ui.adsabs.harvard.edu/abs/2005Natur.435..629S} {435, 629}

\bibitem[\protect\citeauthoryear{{Steidel}, {Dickinson}  \&
  {Persson}}{{Steidel} et~al.}{1994}]{Steidel94}
{Steidel} C.~C.,  {Dickinson} M.,   {Persson} S.~E.,  1994, \mn@doi [\apjl]
  {10.1086/187686}, \href
  {https://ui.adsabs.harvard.edu/abs/1994ApJ...437L..75S} {437, L75}

\bibitem[\protect\citeauthoryear{{Steidel}, {Erb}, {Shapley}, {Pettini},
  {Reddy}, {Bogosavljevi{\'c}}, {Rudie}  \& {Rakic}}{{Steidel}
  et~al.}{2010}]{Steidel10}
{Steidel} C.~C.,  {Erb} D.~K.,  {Shapley} A.~E.,  {Pettini} M.,  {Reddy} N.,
  {Bogosavljevi{\'c}} M.,  {Rudie} G.~C.,   {Rakic} O.,  2010, \mn@doi [\apj]
  {10.1088/0004-637X/717/1/289}, \href
  {https://ui.adsabs.harvard.edu/abs/2010ApJ...717..289S} {717, 289}

\bibitem[\protect\citeauthoryear{{Stocke}, {Keeney}, {Danforth}, {Shull},
  {Froning}, {Green}, {Penton}  \& {Savage}}{{Stocke} et~al.}{2013}]{Stocke13}
{Stocke} J.~T.,  {Keeney} B.~A.,  {Danforth} C.~W.,  {Shull} J.~M.,  {Froning}
  C.~S.,  {Green} J.~C.,  {Penton} S.~V.,   {Savage} B.~D.,  2013, \mn@doi
  [\apj] {10.1088/0004-637X/763/2/148}, \href
  {https://ui.adsabs.harvard.edu/abs/2013ApJ...763..148S} {763, 148}

\bibitem[\protect\citeauthoryear{{Taylor}, {Minchin}, {Herbst}, {Davies},
  {Rodriguez}  \& {Vazquez}}{{Taylor} et~al.}{2014}]{Taylor14}
{Taylor} R.,  {Minchin} R.~F.,  {Herbst} H.,  {Davies} J.~I.,  {Rodriguez} R.,
   {Vazquez} C.,  2014, \mn@doi [\mnras] {10.1093/mnras/stu1305}, \href
  {https://ui.adsabs.harvard.edu/abs/2014MNRAS.443.2634T} {443, 2634}

\bibitem[\protect\citeauthoryear{{Tejos} et~al.,}{{Tejos}
  et~al.}{2014}]{Tejos14}
{Tejos} N.,  et~al., 2014, \mn@doi [\mnras] {10.1093/mnras/stt1844}, \href
  {https://ui.adsabs.harvard.edu/abs/2014MNRAS.437.2017T} {437, 2017}

\bibitem[\protect\citeauthoryear{{Tumlinson} et~al.,}{{Tumlinson}
  et~al.}{2013}]{Tumlinson13}
{Tumlinson} J.,  et~al., 2013, \mn@doi [\apj] {10.1088/0004-637X/777/1/59},
  \href {https://ui.adsabs.harvard.edu/abs/2013ApJ...777...59T} {777, 59}

\bibitem[\protect\citeauthoryear{{Tumlinson}, {Peeples}  \& {Werk}}{{Tumlinson}
  et~al.}{2017}]{Tumlinson17}
{Tumlinson} J.,  {Peeples} M.~S.,   {Werk} J.~K.,  2017, \mn@doi [\araa]
  {10.1146/annurev-astro-091916-055240}, \href
  {https://ui.adsabs.harvard.edu/abs/2017ARA%26A..55..389T} {55, 389}

\bibitem[\protect\citeauthoryear{{Tummuangpak}, {Bielby}, {Shanks}, {Theuns},
  {Crighton}, {Francke}  \& {Infante}}{{Tummuangpak}
  et~al.}{2014}]{Tummuangpak14}
{Tummuangpak} P.,  {Bielby} R.~M.,  {Shanks} T.,  {Theuns} T.,  {Crighton}
  N.~H.~M.,  {Francke} H.,   {Infante} L.,  2014, \mn@doi [\mnras]
  {10.1093/mnras/stu828}, \href
  {https://ui.adsabs.harvard.edu/abs/2014MNRAS.442.2094T} {442, 2094}

\bibitem[\protect\citeauthoryear{{Turner}, {Schaye}, {Steidel}, {Rudie}  \&
  {Strom}}{{Turner} et~al.}{2014}]{Turner14}
{Turner} M.~L.,  {Schaye} J.,  {Steidel} C.~C.,  {Rudie} G.~C.,   {Strom}
  A.~L.,  2014, \mn@doi [\mnras] {10.1093/mnras/stu1801}, \href
  {https://ui.adsabs.harvard.edu/abs/2014MNRAS.445..794T} {445, 794}

\bibitem[\protect\citeauthoryear{{Turner} et~al.,}{{Turner}
  et~al.}{2017}]{Turner17}
{Turner} O.~J.,  et~al., 2017, \mn@doi [\mnras] {10.1093/mnras/stx1366}, \href
  {http://adsabs.harvard.edu/abs/2017MNRAS.471.1280T} {471, 1280}

\bibitem[\protect\citeauthoryear{{Weilbacher} \& {et al.}}{{Weilbacher} \& {et
  al.}}{2014}]{Weilbacher14}
{Weilbacher} P.~M.,  {et al.} 2014, in {Manset} N.,  {Forshay} P.,  eds,  Astr.
  Soc. of the Pacific Vol. 485, Astronomical Data Analysis Software and Systems
  XXIII. p.~451 (\mn@eprint {arXiv} {1507.00034})

\bibitem[\protect\citeauthoryear{{Werk} et~al.,}{{Werk} et~al.}{2014}]{Werk14}
{Werk} J.~K.,  et~al., 2014, \mn@doi [\apj] {10.1088/0004-637X/792/1/8}, \href
  {https://ui.adsabs.harvard.edu/abs/2014ApJ...792....8W} {792, 8}

\bibitem[\protect\citeauthoryear{{Werk} et~al.,}{{Werk} et~al.}{2016}]{Werk16}
{Werk} J.~K.,  et~al., 2016, \mn@doi [\apj] {10.3847/1538-4357/833/1/54}, \href
  {https://ui.adsabs.harvard.edu/abs/2016ApJ...833...54W} {833, 54}

\bibitem[\protect\citeauthoryear{{Werner} et~al.,}{{Werner}
  et~al.}{2004}]{Werner04}
{Werner} M.~W.,  et~al., 2004, \mn@doi [\apjs] {10.1086/422992}, \href
  {http://adsabs.harvard.edu/abs/2004ApJS..154....1W} {154, 1}

\bibitem[\protect\citeauthoryear{{Wetzel}, {Tinker}  \& {Conroy}}{{Wetzel}
  et~al.}{2012}]{Wetzel12}
{Wetzel} A.~R.,  {Tinker} J.~L.,   {Conroy} C.,  2012, \mn@doi [\mnras]
  {10.1111/j.1365-2966.2012.21188.x}, \href
  {http://adsabs.harvard.edu/abs/2012MNRAS.424..232W} {424, 232}

\bibitem[\protect\citeauthoryear{{White} \& {Rees}}{{White} \&
  {Rees}}{1978}]{White78}
{White} S.~D.~M.,  {Rees} M.~J.,  1978, \mn@doi [\mnras]
  {10.1093/mnras/183.3.341}, \href
  {http://adsabs.harvard.edu/abs/1978MNRAS.183..341W} {183, 341}

\bibitem[\protect\citeauthoryear{{Whiting}, {Webster}  \& {Francis}}{{Whiting}
  et~al.}{2006}]{Whiting06}
{Whiting} M.~T.,  {Webster} R.,   {Francis} P.,  2006, \mn@doi [\mnras]
  {10.1111/j.1365-2966.2006.10101.x}, \href
  {https://ui.adsabs.harvard.edu/abs/2006MNRAS.368..341W} {368, 341}

\bibitem[\protect\citeauthoryear{{Wuyts} et~al.,}{{Wuyts}
  et~al.}{2011}]{Wuyts11a}
{Wuyts} S.,  et~al., 2011, \mn@doi [\apj] {10.1088/0004-637X/738/1/106}, \href
  {http://adsabs.harvard.edu/abs/2011ApJ...738..106W} {738, 106}

\bibitem[\protect\citeauthoryear{{Yang}, {Mo}, {van den Bosch}, {Pasquali},
  {Li}  \& {Barden}}{{Yang} et~al.}{2007}]{Yang07}
{Yang} X.,  {Mo} H.~J.,  {van den Bosch} F.~C.,  {Pasquali} A.,  {Li} C.,
  {Barden} M.,  2007, \mn@doi [\apj] {10.1086/522027}, \href
  {http://adsabs.harvard.edu/abs/2007ApJ...671..153Y} {671, 153}

\bibitem[\protect\citeauthoryear{{York} et~al.,}{{York} et~al.}{2000}]{York00}
{York} D.~G.,  et~al., 2000, \mn@doi [\aj] {10.1086/301513}, \href
  {http://adsabs.harvard.edu/abs/2000AJ....120.1579Y} {120, 1579}

\bibitem[\protect\citeauthoryear{{van de Voort}, {Schaye}, {Booth}, {Haas}  \&
  {Dalla Vecchia}}{{van de Voort} et~al.}{2011}]{van-de-Voort11}
{van de Voort} F.,  {Schaye} J.,  {Booth} C.~M.,  {Haas} M.~R.,   {Dalla
  Vecchia} C.,  2011, \mn@doi [\mnras] {10.1111/j.1365-2966.2011.18565.x},
  \href {https://ui.adsabs.harvard.edu/abs/2011MNRAS.414.2458V} {414, 2458}

\bibitem[\protect\citeauthoryear{{van der Wel} et~al.,}{{van der Wel}
  et~al.}{2014}]{van-der-Wel14}
{van der Wel} A.,  et~al., 2014, \mn@doi [\apj] {10.1088/0004-637X/788/1/28},
  \href {http://adsabs.harvard.edu/abs/2014ApJ...788...28V} {788, 28}

\makeatother
\end{thebibliography}

%%%%%%%%%%%%%%%%%%%%%%%%%%%%%%%%%%%%%%%%%%%%%%%%%%

%%%%%%%%%%%%%%%%% APPENDICES %%%%%%%%%%%%%%%%%%%%%

%\appendix

%\section{Appendix}

%As stated.

%%%%%%%%%%%%%%%%%%%%%%%%%%%%%%%%%%%%%%%%%%%%%%%%%%

% Don't change these lines
\bsp % typesetting comment
\label{lastpage}
\end{document}